\documentclass[12pt]{iopart}

\usepackage{iopams}  
\usepackage{psfrag}
\usepackage{upgreek}
\usepackage{amssymb,amscd,amsthm}

\newcommand{\bex}{\begin{exa}}
\newcommand{\eex}{\end{exa}}
\newcommand{\bd}{\begin{Def}}
\newcommand{\ed}{\end{Def}}
\newcommand{\bt}{\begin{theorem}}
\newcommand{\et}{\end{theorem}}
\newcommand{\bl}{\begin{lemma}}
\newcommand{\el}{\end{lemma}}
\newcommand{\be}{\begin{equation}}
\newcommand{\ee}{\end{equation}}
\newcommand{\bea}{\begin{eqnarray}}
\newcommand{\eea}{\end{eqnarray}}

\newcommand{\adots}{\mathinner{\mkern2mu\raise1pt\hbox{.}\mkern2mu
\raise4pt\hbox{.}\mkern2mu\raise7pt\hbox{.}\mkern1mu}}

\newcommand{\beq}{\begin{equation}}
\newcommand{\eeq}{\end{equation}}
\newcommand{\bear}{\begin{array}}
\newcommand{\eear}{\end{array}}
\newcommand\la{{\lambda}}

\newcommand\al{{\alpha}}

\newcommand\om{{\omega}}

\newcommand\dd{\mathrm{d}}
\newtheorem{thm}{Theorem}[section]

\newtheorem{propn}[thm]{Proposition}

\newtheorem{rem}[thm]{Remark}

\newtheorem{exa}[thm]{Example}

\newtheorem{conje}[thm]{Conjecture}

\newtheorem{defn}[thm]{Definition}
\newtheorem{lem}[thm]{Lemma}

\newenvironment{prf}{\trivlist \item [\hskip
\labelsep {\bf Proof:}]\ignorespaces}{\qed \endtrivlist}

\theoremstyle{remark}

\newcommand{\ups}{{\upsilon}}

\newcommand{\Q}{{\mathbb Q}}
\newcommand{\Z}{{\mathbb Z}}
\newcommand{\C}{{\mathbb C}}

\newcommand{\pro}{{\mathrm p}}

\begin{document}

\title[Discrete Painlev\'e equations from Y-systems]{Discrete Painlev\'e equations from Y-systems}


\author[A.N.W.\ Hone]{Andrew N.W. Hone}
\address{School of Mathematics,
Statistics and Actuarial Science, University of
Kent, Canterbury CT2 7NF, UK }
\ead{A.N.W.Hone@kent.ac.uk}

\author[R.\ Inoue]{Rei Inoue}
\address{
Department of Mathematics and Informatics, Faculty of Science, 
Chiba University
Inage, Chiba 263-8522, Japan}
\ead{reiiy@math.s.chiba-u.ac.jp}

\begin{abstract} 
We consider T-systems and Y-systems arising 
from cluster mutations applied to quivers that have the property of being periodic under a sequence of 
mutations. 
The corresponding nonlinear recurrences for cluster variables (coefficient-free T-systems) were described 
in the work of Fordy and Marsh, who completely classified  all such quivers in the case of period 1, and 
characterized them in terms of the
skew-symmetric exchange matrix $B$ that defines the quiver. A broader notion of periodicity in general 
cluster algebras 
was introduced  by Nakanishi, who also described the corresponding Y-systems, and T-systems 
with coefficients.  

A result of Fomin and Zelevinsky says that the coefficient-free T-system provides a solution of the Y-system. 
In this paper, we show that in general there is a
discrepancy  between these two systems, in the sense that the solution of the former does not 
correspond to the general solution of the  latter. 
This discrepancy is removed by introducing additional non-autonomous coefficients into the 
T-system. In particular, we focus on the period 1 case and show that,  when the exchange matrix $B$ is degenerate,
discrete Painlev\'e equations can arise from this construction. 
\end{abstract}

\maketitle

\section{Introduction}

The theory of cluster algebras, introduced by  
Fomin and Zelevinsky more than ten years ago \cite{fz1}, 
has found a wide range of connections 
with different parts of 
mathematics and theoretical physics, 
especially in  Lie theory,  Teichm\"uller theory, Poisson geometry,
discrete integrable systems and string theory.  
Some of the  inspiration for the development of cluster algebras resulted 
from observations of Somos and 
others \cite{gale}, concerning the \textit{Laurent phenomenon} for nonlinear recurrences 
of the form 
\beq 
\label{recf} 
x_{n+N} \, x_n = F(x_{n+1}, \ldots ,  x_{n+N-1}). 
\eeq 
For certain special choices of $F$, a polynomial in $N-1$ variables, 
all iterates are Laurent polynomials 
in the initial data with integer coefficients. In particular, this means that if all 
$N$ initial values are chosen to be 1, then an integer sequence is produced. 
A very well-known example is the Somos-4 recurrence  
\beq\label{somos4}
x_{n+4}\, x_n=x_{n+3}\, x_{n+1}+x_{n+2}^2;  
\eeq
the integer sequence beginning $1,1,1,1,2,3,7,23,59,314,1529,8209,83313,\ldots$ 
(for details, see {\tt http://oeis.org/A006720}) is generated by (\ref{somos4}) 
starting from four initial 1s,  
while if the initial data $x_1,x_2,x_3,x_4$ are viewed 
as variables then the iterates $x_n$ belong to the Laurent polynomial ring 
$\Z[x_1^{\pm 1}, x_2^{\pm 1}, x_3^{\pm 1}, x_4^{\pm 1}]$. 

In general, the clusters in a cluster algebra of rank $N$ are generated by sequences of cluster mutations, 
defined by exchange relations of the form 
$\tilde{x} x =M_1 +M_2$, where (as in (\ref{somos4}) above) the right-hand side 
is a sum of two monomials $M_1,M_2$, and it is known that all clusters consist of Laurent 
polynomials in the cluster variables ${\bf x} = (x_1,x_2,\ldots, x_N)$ 
from any initial seed \cite{fz1}. However, the Laurent phenomenon can arise in a broader context, 
including recurrences (\ref{recf}) where $F$ is a sum of more than two monomials \cite{fz}, 
which has led to the development of so-called LP algebras \cite{lp}. 
 
Cluster algebras and the Laurent phenomenon appear in connection with integrable 
systems in various different ways, both at the classical and quantum level, 
and for both continuous and discrete systems.  On the one hand, the Hirota-Miwa  equation 
(also known as the discrete Hirota equation, the discrete  KP equation, or the octahedron recurrence) 
is an example of a partial difference equation with the Laurent 
property \cite{fz}; it appears as an identity for 
transfer matrix elements in quantum integrable models \cite{zabrodin}, 
and with particular boundary conditions it is obtained from sequences of cluster mutations 
in a cluster algebra of infinite rank \cite{qsystems}. On the other hand, for 
the continuous KP equation there is a cluster algebra structure in the  
graphs describing the combinatorics of the interaction of solitons \cite{yuji_lauren};  
the latter connection can be understood from the fact that the tau-functions 
of KP soliton solutions are encoded by points in a finite-dimensional 
Grassmannian, whose cluster algebra structure was explained by Scott \cite{scott}.  
Examples such as these suggest that 
tau-functions of integrable systems should admit an interpretation 
as cluster variables in a suitable cluster algebra.   
This interpretation can be very useful: 
there are natural Poisson/symplectic structures associated with 
cluster algebras  \cite{fockgon,gsv,gsvduke}, which may be 
used to obtain Poisson brackets for discrete integrable systems, 
both partial difference equations \cite{inoue} and integrable maps \cite{dkdv}.  

There are many other connections between cluster algebras and integrability. 
The original Y-systems of Zamolodchikov \cite{zam}, 
which are functional relations arising from the thermodynamic Bethe ansatz for 
certain integrable quantum field theories, were the prototype for the 
dynamics of coefficients in cluster algebras, and periodicities in Y-systems 
continue to be the subject of active research \cite{iikns}. In the context of cluster algebras, 
there are also various examples 
of discrete integrable systems that are linearizable, in the sense that the variables  
satisfy linear recurrences. 
In particular, 
Fordy and Marsh considered cluster algebras obtained from quivers which are 
periodic under sequences of mutations \cite{fordy_marsh}, and showed that in certain cases, 
corresponding to affine A-type Dynkin quivers, linear relations hold between the cluster variables; 
for types A and D this was found independently in the context of frieze patterns \cite{assem}, and 
linear relations for all affine Dynkin types were proved in \cite{keller_scher}. Starting with  
results in \cite{honelaur},  Poisson-commuting first 
integrals were constructed for affine A-type systems, as well as for  
other linearizable systems obtained from cluster algebras with periodicity  \cite{fordy_rec,fh}.  
For certain linearizable systems, first integrals are also 
known in the noncommutative case \cite{difrancesco}.
Outside the context of cluster algebras, there are other linearizable systems with the Laurent 
property (see e.g. \cite{heidhogan08,numberpoly}), within the 
LP algebra framework 
\cite{alman,honeward}.

In this paper we point out  
a new link between cluster algebras and integrable systems, by showing how 
certain discrete Painlev\'e equations, as well as their higher-order analogues, 
can be constructed from Y-systems. 
Our starting point is Nakanishi's 
construction of generalized T-systems and Y-systems associated with 
cluster algebras with periodicity under mutations \cite{nakanishi}, which 
is based on a broader 
notion of periodicity than that considered in 
\cite{fordy_marsh}. 
Nevertheless, for the most part we concentrate on the simplest case of Y-systems arising from quivers 
with period 1 under mutation;  such quivers were completely classified by the authors of \cite{fordy_marsh}, who  
showed that they generate  
recurrences (coefficient-free T-systems) of the general form 
\beq\label{arec} 
x_{n+N}\, x_n = \prod_{a_j\geq 0}x_{n+j}^{a_j}+ \prod_{a_j\leq 0}x_{n+j}^{-a_j}, 
\eeq 
where the indices in each product lie in the range $1\leq j\leq N-1$, with the 
exponents $(a_1,...,a_{N-1})$ forming an integer $(N-1)$-tuple which is   
palindromic, so that $a_j = a_{N-j}$.  
We discuss the Y-systems associated with (\ref{arec}), which have the 
form 
\beq\label{ayrec} 
{y}_{n+N}\, {y}_n = \frac{\prod_{a_j\leq 0}\, (1+{y}_{n+j})^{-a_j}} {\prod_{a_j\geq 0}\, (1+{y}_{n+j}^{-1})^{a_j}}, 
\eeq 
and describe the relation between the solutions of (\ref{arec}) 
and (\ref{ayrec}) as explicitly as possible:     
in general, the solution space of (\ref{arec}) 
corresponds to a space of lower dimension 
within the solutions of (\ref{ayrec}). In particular, when 
the recurrence (\ref{arec}) is of Somos-$N$ type, i.e. 
\beq \label{somosN}
x_{n+N}\, x_n = x_{n+N-p}\, x_{n+p} + x_{n+N-q}\, x_{n+q}
\eeq  
with $N\geq 2$ and $1\leq p <q\leq
\lfloor{N/2}\rfloor$, 
then the general solution of the associated Y-system 
corresponds to a non-autonomous difference equation of q-Painlev\'{e} type.

\subsection{Outline of the paper}

In section 2 we consider the Somos-4 recurrence and its connections 
with various topics in mathematical physics 
(QRT maps and dimer models in particular), before explaining how the 
corresponding Y-system is related to a q-Painlev\'e I equation. 
Section 3 is concerned 
with coefficient-free cluster algebras and cluster mutation-periodicity, and especially the 
period 1 case as classified in   \cite{fordy_marsh}. Following \cite{fh}, we describe 
how symplectic maps arise in the latter context, and introduce a special family 
of recurrences (each member of which is equivalent to iteration of a symplectic map) 
called U-systems (Definition~\ref{udef}). 
The fourth section gives a very brief description of cluster algebras with coefficients, setting 
the scene for 
the 
more general notion of periodicity in \cite{nakanishi}, and the Y-systems 
and T-systems that arise in that context. A modified version of T-systems is introduced, called T$_z$-systems,  
together with so-called Z-systems, given by equations (\ref{mtsysi}) and (\ref{zsysi}) respectively. 
The period 1 case is considered in more 
detail, examining the precise relation between T$_z$-systems and Y-systems in that case.       
Section 5 is devoted to some specific examples of T$_z$-systems in the period 1 case, 
in particular showing how discrete Painlev\'e equations, and their higher-order analogues, 
appear from this construction. We make a few conclusions in section 6, and the proof of a technical 
result (Proposition \ref{palilem}) 
is reserved for an appendix. 

A few  of our results appeared in unpublished sections of {\tt arXiv:1207.6072v1}, the original preprint 
version of \cite{fh}.

\section{Motivating example: Somos-4 and q-Painlev\'{e} I}
\setcounter{equation}{0}

It was noted in \cite{inoue} that for the case of cluster algebras associated with Lotka-Volterra systems
the solution of the T-system  corresponds to 
a system of lower order than the Y-system. 
Nakanishi pointed out to the first author 
that the same phenomenon occurs for the Somos-4 recurrence (\ref{somos4}), which serves 
as motivation for  the 
rest of the paper. 


\subsection{
Symplectic coordinates and Liouville integrability} 

The Somos-4 recurrence (\ref{somos4}) 
is the simplest three-term bilinear recurrence of the form (\ref{somosN}). In can be 
obtained as a particular 
reduction of the discrete Hirota equation, which leads to a Lax pair 
and associated spectral curve for it \cite{fh}. In subsequent sections we shall explain 
how it arises in the context of cluster algebras as a (generalized) T-system, 
but here we just describe  particular properties of this equation.  

Due to its bilinear nature, 
it is clear that the equation  (\ref{somos4})  
is invariant under the two-dimensional group of scaling transformations 
\beq\label{gauge} 
x_n \longrightarrow \lambda \, \mu^n \, x_n, \qquad (\lambda ,\mu )\in (\C^*)^2.  
\eeq 
This is a simple example of the gauge transformations that appear in Hirota's 
theory of tau-functions for soliton equations \cite{hir}. 
The ``physical variables'' are the gauge-invariant quantities 
\beq\label{yvars} 
Y_n = \frac{x_nx_{n+2}}{x_{n+1}^2}, 
\eeq 
which satisfy the second-order recurrence 
\beq \label{s4qrt} 
Y_{n+2}\, Y_n = \frac{Y_{n+1} + 1}{Y_{n+1}^2}. 
\eeq 
The above recurrence for $Y_n$ is solved in elliptic functions, which leads 
to the explicit analytic solution of (\ref{somos4}), as 
in \cite{honeblms}. 

There is another way to obtain the variables (\ref{yvars}). 
Iterations 
of the Somos-4 recurrence preserve a degenerate quadratic Poisson bracket 
in the variables $x_n$ \cite{honelaur}, but of 
more importance 
for the dynamics   is 
the presymplectic form
\beq   
\label{7sympm1is1} 
\bear{rcl}
\omega &   =  &
  -\left( \frac{\dd x_1 \wedge \dd x_2}{x_1x_2}+\frac{\dd x_1 \wedge \dd x_4}{x_1x_4}+
\frac{\dd x_3 \wedge \dd x_4}{x_3x_4}\right)    
\\ && +2 \left(\frac{\dd x_1 \wedge \dd x_3}{x_1x_3}+ \frac{\dd x_2 \wedge \dd x_4}{x_2x_4}\right)
-3 \, \frac{\dd x_2 \wedge \dd x_3}{x_2x_3}, 
\eear
\eeq 
which is also preserved by the map $\varphi: \, (x_1,x_2,x_3,x_4)\mapsto (x_2,x_3,x_4,x_5)$. 
This closed two-form in four dimensions is degenerate, but  $(Y_1,Y_2)$ are coordinates on the space of leaves 
of the null foliation for $\om$, and on this space $\om$  reduces to the symplectic 
form 
$$ 
\hat{\om} = (Y_1Y_2)^{-1} \dd Y_2\wedge \dd Y_1 , 
$$
which gives a log-canonical Poisson bracket 
between these coordinates: 
\beq \label{s4bra}  
\{Y_1,Y_2\} = Y_1Y_2 .
\eeq

Iterates of (\ref{somos4}) produce 
iterates of (\ref{s4qrt}), and in 
the $(Y_1,Y_2)$-plane this 
corresponds to iteration of the symplectic map
\beq \label{s4map}
\hat{\varphi}: \qquad
\left(\begin{array}{c}
Y_1 \\
Y_2
\end{array} \right)
\longmapsto  
\left(\begin{array}{c}
\tilde{Y}_1 \\
\tilde{Y}_2
\end{array} \right)
=
 \left(\begin{array}{c}
Y_2 \\
(Y_2+1)/(Y_1Y_2^2)
\end{array} \right),
\eeq
which is of QRT type \cite{qrt1}. A geometrical construction 
of this map, which goes back to Chasles (or even Euler)  \cite{veselov}, 
is achieved by starting from the following family of biquadratic 
curves: 
\beq \label{biq} 
(Y_1Y_2)^2 -H\,Y_1Y_2 +Y_1+Y_2 +1=0.    
\eeq 
For fixed $H$, the above curve admits the pair of involutions 
$$ 
\iota_1: \, Y_1\leftrightarrow Y_2, 
\qquad \iota_2: \, (Y_1,Y_2) \mapsto (Y_1^\dagger ,Y_2), 
$$ 
where the ``horizontal switch'' $\iota_2$ 
gives the conjugate point $(Y_1^\dagger,Y_2)$ obtained 
by intersecting the curve with a horizontal line. 
The formula for the product of the roots of 
(\ref{biq}), viewed as a quadratic in $Y_1$, 
gives the expression  
$Y_1^\dagger = (Y_2+1)/(Y_1Y_2^2)$, from which 
it is clear that the map (\ref{s4map}) is the composition $\hat{\varphi}=\iota_1\cdot\iota_2$.  
Moreover, 
for generic $H$ the curve (\ref{biq}) has genus one, and all 
formulae are independent of this 
parameter, so solving (\ref{biq}) for $H$ gives a first integral; hence  
the map $\hat\varphi$ is an  integrable system with one degree of freedom, in the Liouville sense 
\cite{maeda,veselov}, and the level sets of the Hamiltonian $H$ are elliptic curves. 
By performing a sequence of blowups at singularities, this map lifts to a morphism 
of a smooth elliptic surface \cite{tsuda_elliptic,QRT}.   


\begin{rem} \label{s4dilog} 
{\em 
In terms of the Rogers dilogarithm function 
$$ 
L(\upzeta)  =  -\frac{1}{2} \int_0^\upzeta \left(\frac{\log (1-y)}{y} + \frac{\log y}{(1-y)} \right) \, \dd y \\ 
 =  Li_2(\upzeta ) +  \frac{1}{2}\log\upzeta \, \log (1-\upzeta ), 
$$ 
the symplectic map (\ref{s4map}) 
has the generating function 
$$ 
\hat{G}(Y_2,\tilde{Y}_2)=L\left(\frac{1}{1+Y_2}\right)+\log Y_2\log \tilde{Y}_2 + (\log Y_2)^2 -\frac{1}{2}\log Y_2\log(1+Y_2), 
$$ 
such that $\log\tilde{Y}_1\, \dd\log \tilde{Y}_2 - \log Y_1\, \dd \log Y_2 =\dd \hat{G}$. 
We present dilogarithmic 
generating functions for 
T-systems coming from periodic quivers in Lemma \ref{dilog} below; 
the particular function $\hat G$ above follows from a particular case of this result. 
It is interesting to note that dilogarithms also appear in the Lagrangians 
for integrable lattice equations \cite{lobb}, 
while 
the main subject of \cite{nakanishi} is the 
identities for the Rogers dilogarithm which are 
associated with the Y-systems of general cluster 
algebras with periodicity.
}  
\end{rem}

\subsection{Dimer models and relativistic Toda lattices} 

Recently, Goncharov and Kenyon have shown that dimer models on a torus give rise to certain quantum integrable systems,
referred to as cluster integrable systems  \cite{gk}. Each of these systems has a classical limit 
whose Lax matrix is the Kasteleyn matrix of the dimer model, 
with a  combinatorial  construction giving  the Poisson-commuting Hamiltonians 
as  sums of weighted dimer covers.   
The partition function of the dimer model is the spectral curve of the classical system. 
Furthermore, these 
systems have discrete symmetries given by cluster exchange relations, which are  bilinear (Somos-type) equations.
Thus the Somos recurrences of the form (\ref{somosN}) can be understood as discrete integrable systems
arising as discrete symmetries (or B\"acklund transformations \cite{kuskly}) of these continuous systems.
In a further development, Eager et al. \cite{eager} have observed that in certain cases, associated with
$Y^{\pro ,0}$ toric surfaces, the classical cluster integrable systems
arising in this way are equivalent to the relativistic Toda lattices in \cite{ruijs}, 
while the dimer models associated with other $Y^{\pro ,\mathrm{q}}$ geometries 
apparently produce new relativistic Toda systems. 
The Somos-4 recurrence (\ref{somos4}) being discussed here corresponds to  $Y^{2,0}$ (that is, $\pro=2$); in \cite{eager2} 
it is explained that it also arises from $Y^{2 ,1}$ (del Pezzo 1).

The classical cluster integrable systems corresponding to
$Y^{ \pro ,0}$ geometries are equivalent to the relativistic Toda lattices,
which live on a $2\pro$-dimensional phase space with coordinates
$(c_j,d_j)_{j=1}^\pro$ whose non-vanishing Poisson
brackets are given by
\beq\label{ruijsbra} 
\{ c_j,c_{j+1}\}=- c_jc_{j+1}, \quad \{ c_j,d_j\}=c_jd_j, \quad
\{ c_j,d_{j+1}\}=-c_jd_{j+1}
,
\eeq 
where indices are read $\bmod$ $\pro$. The first Hamiltonian is
\beq\label{ruijsh1} 
H_1=\sum_j c_j +d_j,
\eeq 
and there are $\pro -2$ commuting higher Hamiltonians as well as two Casimirs, 
given by $\prod_{j=1}^\pro c_j$ and $\prod_{j=1}^\pro d_j$ for $\pro\neq 2$, so that the system
is integrable with symplectic leaves of dimension $2\pro -2$.
From  \cite{fg} there is a B\"acklund transformation (in the sense of \cite{kuskly}) given by 
\beq\label{bt}
\tilde{c}_j=c_j\, \left(\frac{d_j+c_{j-1}}{d_{j+1}+c_j}\right), \qquad
\tilde{d}_j=d_{j+1}\,\left(\frac{d_j+c_{j-1}}{d_{j+1}+c_j}\right)
\eeq
for $\pro\neq 2$; this is a Poisson map that preserves the Hamiltonians and Casimirs. 

In the case $\pro =2$, the Poisson tensor has rank 2, with $c_1c_2d_1$ and $d_1d_2$ being the two independent
Casimirs.    If we set
$$
Y_1=\sqrt{\frac{c_1d_1}{c_2}}, \qquad Y_2=\sqrt{\frac{c_2d_1}{c_1}},
$$
then (\ref{ruijsbra}) gives the log-canonical bracket (\ref{s4bra}) 
for the $Y_j$, and on the symplectic leaves
\beq\label{sleaves} 
\alpha = \sqrt{c_1c_2d_1}=\mathrm{constant}, \qquad \beta = d_1d_2=\mathrm{constant} 
\eeq 
the Hamiltonian $H_1$ in 
(\ref{ruijsh1}) becomes precisely the first integral for Somos-4,  obtained by solving 
(\ref{biq}) for $H$. For $\pro =2$, the Poisson map (\ref{bt}) must be modified as
$$ 
\tilde{c}_1=\frac{c_2d_1}{c_2+d_2}, \quad 
\tilde{c}_2=c_1, \quad 
\tilde{d}_1=c_2+d_2, \quad 
\tilde{d}_2=\frac{d_1d_2}{c_2+d_2}. 
$$ 
It is easy to verify that, on the level of the symplectic leaves (\ref{sleaves}), the iteration of this map is 
identical to (\ref{s4map}).


\subsection{Somos-4 Y-system and a discrete Painlev\'{e} equation}

The Somos-4 recurrence (\ref{somos4}) is a generalized T-system, in the sense 
of \cite{nakanishi}; it has the form (\ref{arec}), with $N=4$ and the palindromic 
triple $(a_1,a_2,a_3)=(-1,2,-1)$.   The generalized Y-system corresponding to 
Somos-4, of the form (\ref{ayrec}), is  
given by 
\beq \label{s4y} 
y_{n+4}\, {y}_n = \frac{(1+{y}_{n+3})(1+{y}_{n+1})} {(1+{y}_{n+2}^{-1})^2}.  
\eeq   
It turns out that the general solution of this Y-system is related to 
non-autonomous versions of (\ref{somos4}) and (\ref{s4qrt}), which is 
how a discrete  Painlev\'{e} equation appears.

There is an algebraic connection between the solutions of T- and Y-systems, 
which corresponds to a general link between cluster variables (denoted by ${\bf x}$) 
and coefficient variables (denoted by ${\bf y}$) that was presented 
in Proposition 3.9 of \cite{fziv} (cf. Proposition 5.11 in \cite{nakanishi}). 
In the particular case at hand, it means that  
substituting $y_n=Y_n$ 
with  $Y_n$ as in (\ref{yvars})  
gives a solution of the Y-system (\ref{s4y}) whenever $x_n$  
satisfies (\ref{somos4}). 
Yet there is a discrepancy between the Y-system (\ref{s4y}), 
which is a fourth-order recurrence, and (\ref{s4qrt}), which is 
only second-order; although (\ref{somos4}) is fourth-order, the 
gauge symmetry (\ref{gauge}) reduces the dimension by two. Every solution of 
(\ref{s4qrt}) is a solution of  (\ref{s4y}), but 
the converse is not true: the recurrence (\ref{s4y}) 
requires four initial values, so the general solution should 
depend on four arbitrary constants, whereas the general solution 
of (\ref{s4qrt}), which is given explicitly in terms of 
elliptic functions in \cite{honeblms}, depends on only 
two parameters.

In order to better understand the general solution of  (\ref{s4y}), 
consider the quantity 
$$ 
\mathcal{Z}_n:= \frac{{y}_{n+2}\, {y}_{n+1}^2\,{y}_{n}} {1+{y}_{n+1}},  
$$ 
which is defined so 
that $\mathcal{Z}_n=1$ for all $n$ whenever ${y}_n=Y_n$  
satisfies (\ref{s4qrt}). 
In general, the Y-system (\ref{s4y}) holds 
if and only if $\mathcal{Z}_n$ 
satisfies  
\beq \label{doll} 
\frac{ \mathcal{Z}_n\,\mathcal{Z}_{n+2}}{\mathcal{Z}_{n+1}^2}=1, 
\eeq and by taking logarithms this gives a linear difference 
equation which implies that $\log\mathcal{Z}_n$ is a linear function of 
$n$, hence $\mathcal{Z}_n=\beta\, \mathfrak{q}^n$ for constants 
$\beta \, ,\mathfrak{q}\in\C^*$.  Upon rewriting the 
definition of $\mathcal{Z}_n$, the Y-system can be rewitten 
as a second-order non-autonomous recurrence for ${y}_n$.

\begin{propn} \label{dpi} 
The general solution of the Somos-4 Y-system is given by 
a solution of the second-order recurrence 
\beq \label{pi} 
{y}_{n+2}\, {y}_n = \beta\,  \mathfrak{q}^n\, \frac{(1+{y}_{n+1})}{{y}_{n+1}^2}, 
\eeq 
which is a $\mathfrak{q}$-difference analogue of the first 
Painlev\'e equation. 
\end{propn} 

The above statement is obvious from the foregoing 
discussion, apart from the identification 
of (\ref{pi}) as a known example of a discrete Painlev\'e equation. 
To see this, 
one can make 
a gauge transformation of the dependent variable in 
the equation (\ref{pi}), setting ${y}_n=\beta^{-1}\, \alpha_n\, \ups_n$ 
for a suitable function $\alpha_n$, 
chosen 
so that one of the coefficients on the right-hand side of the recurrence 
becomes constant. We obtain 
\beq\label{cpi} 
\ups_{n+2}\, \ups_{n} = \frac{(\alpha_n \, \ups_{n+1} + \beta)} 
{\ups_{n+1}^2}, \qquad \mathrm{where} \quad 
\alpha_{n+2}\, \alpha_{n+1}^2 \, \alpha_n=\beta^4 \, 
\mathfrak{q}^n. 
\eeq 
The above equation for $u_n$ is one of the discrete Painlev\'e I equations 
derived using the singularity confinement method in  \cite{dpi}, where 
it is shown that the non-autonomous coefficient $\al_n$ must  
satisfy $\al_{n+4}\, \al_n = \al_{n+2}^2$ in order for the 
singularitities to be confined. The latter condition 
is a consequence of the second-order relation for $\al_n$ in 
(\ref{cpi}); it implies that this coefficient takes the alternating form  
$\al_n = \al_{\pm}\, \mathfrak{q}_{\pm}^n$ for even/odd $n$. 
To see the link with Painlev\'e differential equations, 
one should take a continuum limit $\ups_n = h^{-2}-U(nh)$ with $s=nh$ 
held fixed as $h\to 0$, and scale $\al_n$ and $\beta$ suitably, 
to obtain the differential equation 
$$ 
\frac{d^2U}{ds^2} = 6U^2 + s, 
$$ 
which is the Painlev\'e I equation. (See 
the last section of \cite{honep} for more details of the continuum limit.)    
The special case $\al_n=\al=\, $constant for all $n$ corresponds to the 
general Somos-4 recurrence with constant coefficients $\al$, $\beta$, which was solved 
analytically in 
\cite{honeblms}, and was obtained from a quiver with frozen nodes in 
\cite{fordy_marsh}.  

\begin{rem} \label{dpibilinear}{\em  
The substitution (\ref{tsubs}) 
gives  a particular solution of the Y-system 
in terms of a solution of the T-system, 
but one can use the same substitution in the general case, without  
the assumption that ${x}_n$ satisfies the coefficient-free T-system (\ref{somos4}). 
Indeed, setting 
${y}_n = {x}_{n}{x}_{n+2}/{x}_{n+1}^2$ in 
(\ref{pi}) implies that  ${x}_n$  is a solution of the 
non-autonomous T-system 
\beq\label{bildpi}  
{x}_{n+4}\, {x}_{n}=\beta\,  \mathfrak{q}^n\,  ({x}_{n+3}\, {x}_{n+1}+ {x}_{n+2}^2),
\eeq 
which is equivalent to one of the bilinear forms of  
discrete Painlev\'e I equations obtained in \cite{dpi}.
The Laurent property is preserved in the presence of 
the non-autonomous coefficients, in the sense 
that the iterates of (\ref{bildpi}) 
belong to the ring $\Z[x_0^{\pm 1},x_1^{\pm 1},x_2^{\pm 1},x_3^{\pm 1},
\mathfrak{q}^{\pm 1},\beta ]$ for all $n$; some related observations appear  
in the unpublished preprint \cite{side6_poster}. 
}
\end{rem}     
%
%
\begin{rem}{\em 
Each iteration of the non-autonomous recurrence (\ref{pi}) 
preserves the  symplectic form 
$\hat\om = ({y}_n \, {y}_{n+1})^{-1} \, \dd {y}_{n+1} \wedge \dd {y}_{n}$, 
which is the same as for the QRT map (\ref{s4qrt}) in the autonomous case. 
} 
\end{rem}
Note that the exponents of $\mathcal{Z}_n$ on the left 
hand side of (\ref{doll}) are the same as those in the substitution 
(\ref{yvars}) 
that gives a solution of the Somos-4 Y-system. In 
section 4 we derive analogous results for other Y-systems 
obtained from cluster algebras with periodicity. 
Our main examples are associated 
with cluster algebras from quivers, which we introduce in the 
next section.

\section{Cluster recurrences and symplectic maps from quivers} \label{torusaction}
\setcounter{equation}{0}

In this section we introduce the basic notions of coefficient-free cluster algebras obtained 
from quivers, before describing cluster mutation-periodic quivers and associated 
(autonomous) recurrence relations and symplectic maps.   

\subsection{Recurrences from periodic quivers} 

In a coefficient-free cluster algebra of rank $N$, 
a   seed $(B,{\bf x})$  consists 
of an {\it exchange matrix } $B=(b_{ij})\in\mathrm{Mat}_{N}(\Z)$, which is 
skew-symmetrizable 
(i.e. there is a positive diagonal integer matrix $D$ such that 
$B^TD=-DB$), and a {\it cluster} ${\bf x}$,  which is an $N$-tuple of cluster 
variables ${\bf x}= (x_1,x_2, \ldots , x_N)$. 
For $j,k \in \mathbb{Z}$ such that $j \leq k$, 
we write $[j,k]$ for $\{j,j+1,\ldots,k-1,k\}$. 
For $k \in [1,N]$, the mutation $\mu_k$ 
gives 
a new seed 
$(\tilde{B},\tilde{{\bf x}})=(\mu_k (B), \mu_k ({\bf x})) $, with 
$\tilde{B}=(\tilde{b}_{ij})$ and $\tilde{{\bf x}}=(\tilde{x}_1,\tilde{x}_2, \ldots , \tilde{x}_N)$, 
where matrix mutation is defined by 
\beq\label{matmut}
\tilde{b}_{ij} =
\left\{
\begin{array}{ll}
-b_{ij}\qquad & \mathrm{if} \quad
i=k\quad  \mathrm{or}\quad j=k, \\
b_{ij} + \frac{1}{2}(|b_{ik}|b_{kj} + b_{ik}|b_{kj}|) & \mathrm{otherwise}, 
\end{array}
\right.
\eeq 
while cluster mutation is given by 
\beq\label{exch}
\tilde{x}_i
=
\left\{
\begin{array}{ll}
\displaystyle{
\frac{ 
\prod_{i \in [1,N]}x_{i}^{[b_{ik}]_+}\,  
+\prod_{i \in [1,N] }x_{i}^{[-b_{ik}]_+}}{x_k}},
\qquad & i = k,
\\[1mm]
x_i, & i \neq k,
\end{array}
\right.
\eeq
where $[b]_+=\max (b,0).$
%
The {\it cluster algebra} ${\cal A} ={\cal A}(B)$ is the algebra over $\Z$   
generated by all of the cluster variables obtained by all possible 
sequences of mutations, and the Laurent property means that 
all cluster variables belong to 
$\Z[{\bf x}^{\pm 1}]:= \Z [x_1^{\pm 1}, \ldots ,x_N^{\pm 1}]$, the ring of Laurent polynomials 
in the initial cluster \cite{fz1}.    

A quiver is a graph consisting of a number of nodes together 
with arrows between the nodes. 
To each quiver $Q$ with $N$ nodes, without 1- or 2-cycles (that is, 
without any paths $i\rightarrow i$ or $i\rightarrow j \rightarrow i$ for $i\neq j$), there corresponds 
a skew-symmetric integer matrix $B\in\mathrm{Mat}_{N}(\Z)$, and vice-versa.  
For any such quiver, one can apply quiver mutation $\mu_k$ at node $k$, 
which  acts as follows: 
(i) reverse 
all arrows in/out of node $k$; (ii) if there are $p$ arrows from 
node $j$ to node $k$, and $q$ arrows from node $k$ to node $\ell$, 
then add $pq$ arrows from node $j$ to node $\ell$; (iii) remove 
any 2-cycles created in step (ii). The latter operation sends 
$Q$ to $\tilde{Q}=\mu_k(Q)$, and is equivalent to the formula  (\ref{matmut}) 
for matrix mutation
in  the skew-symmetric case  $b_{ij}=-b_{ji}$.   

In general, iteration of  (\ref{exch}) cannot be interpreted as a discrete dynamical 
system, because there are $N$ possible directions for mutation at each step, and the matrix mutation 
(\ref{matmut}) alters the exponents that appear in the two monomials on the right hand side 
of  the exchange relation. However, in \cite{fordy_marsh}, for the skew-symmetric case
where the exchange matrix is associated with a quiver, 
 $B$ was defined to be {\it  cluster mutation-periodic }  
with period $m$ if (for a suitable labelling of indices) $\mu_m \cdot \mu_{m-1}\cdot \ldots \cdot \mu_1 (B) = \rho^m (B)$, 
where $\rho$ is a cyclic permutation.  
In this setting, the {\it cluster map} $\varphi = \rho^{-m}\cdot \mu_m \cdot \mu_{m-1}\cdot \ldots \cdot \mu_1$ 
acts as the identity on $B$, but generically  ${\bf x}\mapsto \varphi ({\bf x})$ has a non-trivial 
action on the cluster, and generates an infinite sequence of cluster variables; thus one has 
discrete dynamics corresponding to mutations in a special sequence of directions.

For the case of period $m=1$, cluster mutation-periodicity for a quiver $Q$ means 
that (with appropriately labelled nodes) the action of mutation $\mu_1$ at node 1 
on $Q$ has the same effect as the action of $\rho: \, (1,2,3,\ldots,N)\mapsto (N,1,2,\ldots,N-1)$, such that the number of arrows
from $j$ to $k$ in $Q$ is the same as the number of arrows from $\rho^{-1}(j)$ to $\rho^{-1}(k)$ in $\rho (Q)$. 
Then the action of $\varphi =\rho^{-1}\cdot \mu_1$ on the cluster ${\bf x}$ takes the form 
of the  birational map 
\beq \label{bir}
\varphi: \quad 
(x_1, 
x_2, 
\ldots , 
x_{N-1}, 
x_{N}) 
\longmapsto 
(
x_2 ,
x_3 , 
\ldots , 
x_{N} , 
x_{N+1}
) , 
\eeq 
where
$$ 
x_{N+1}=\frac{
\prod_{j=1}^{N-1} x_{j+1}^{[b_{1,j+1}]_+} + \prod_{j=1}^{N-1} x_{j+1}^{[-b_{1,j+1}]_+}
 }{x_1}.
$$ 
Iterating this map 
is equivalent to the iteration of a single scalar recurrence relation, that is  
\beq\label{crec}
x_{n+N}x_n = \prod_{j=1}^{N-1} x_{n+j}^{[b_{1,j+1}]_+} + \prod_{j=1}^{N-1} x_{n+j}^{[-b_{1,j+1}]_+}, \qquad n=1,2,3,\ldots.
\eeq
\bex \label{s4bper} 
{ \em 
The exchange matrix 
\beq\label{s4bmat}
B = \left(\begin{array}{cccc} 0 & -1 & 2 & -1 \\ 
                                               1 & 0 & -3 & 2 \\ 
                                                -2 & 3 & 0 & -1 \\ 
                                               1 & -2 & 1 & 0 \end{array}\right)  , 
\eeq 
with $N=4$, 
satisfies  $\mu_1( B) = \rho (B)$, 
and the map 
$\varphi = \rho^{-1} \cdot \mu_1$ acting on the cluster ${\bf x}=(x_1,x_2,x_3,x_4)$ is 
equivalent to an iteration of the Somos-4 recurrence (\ref{somos4}), which 
has the form (\ref{crec}). 
} 
\eex 

A complete 
classification of period 1 quivers is given in \cite{fordy_marsh}. 
Cluster mutation-periodicity with period 1 holds if and only if  
the matrix elements of $B$ satisfy the relations
\beq\label{reln1}
b_{j,N}=b_{1,j+1}, \quad j\in [1,N-1], 
\eeq
and
\beq\label{reln2}
b_{j+1,k+1}=b_{j,k}+b_{1,j+1} [-b_{1,k+1}]_+  - b_{1,k+1} [-b_{1,j+1}]_+  ,
\quad j,k \in [1,N-1]. 
\eeq
The above formulae 
entail that a matrix $B$ associated with 
a period 1 cluster mutation-periodic quiver is completely determined by 
the elements in its first row, where the integers 
$a_j=b_{1,j+1}$ for $j=1,\ldots ,N-1$ form a 
palindromic integer $(N-1)$-tuple ${\bf a}=(a_1,\ldots ,a_{N-1})$, i.e. 
$a_j=a_{N-j}$ (cf. Theorem 6.1 in \cite{fordy_marsh}).
Moreover, apart from being skew-symmetric,  such a matrix $B$ is also symmetric about the skew diagonal, 
i.e. $b_{jk}=b_{N-k+1,N-j+1}$. 
Hence each recurrence of the form (\ref{arec}) with 
palindromic exponents $a_j$ corresponds to a matrix $B$ of this kind, 
and conversely.

\subsection{Symplectic forms for cluster maps} 

For a skew-symmetric integer matrix $B$,  
it was shown in \cite{gsvduke} (see also \cite{fockgon}) 
that the two-form
\beq \label{omega}
\om =\sum_{j<k} \frac{b_{jk}}{x_jx_k}\dd x_j\wedge \dd x_k   
\eeq
transforms covariantly with respect to cluster mutations $\mu_i$. 
In the special case of cluster mutation-periodicity with period 1, 
it turns out that (\ref{omega}) is invariant under the action of 
the cluster map (\ref{bir}). 
Upon introducing the one-form 
$$\vartheta =\sum_{j<k}b_{jk}\, z_j\, \dd z_k, 
\qquad \mathrm{with} \qquad z_j =\log x_j, 
$$
so that 
$
\om=\dd\vartheta, 
$
the fact that $\varphi^* \om =\om$ can be seen from the existence 
of a generating function for $\varphi$, given in terms of  
the Rogers dilogarithm function
(as in Remark \ref{s4dilog}).  
For the rest of this section we assume that $B$ 
is a skew-symmetric integer matrix corresponding to a cluster mutation-periodic quiver with period 1.

\begin{propn}\label{dilog} 
The map (\ref{bir})  has the generating function $G=G_0 + G_L$, such that 
$\varphi^*\vartheta - \vartheta = \dd G$, where 
$$ 
G_0 = \sum_{1\leq j<k\leq N-1} [-b_{1,j+1}]_+\, b_{1,k+1} \, z_{j+1}z_{k+1} 
+\sum_{j=1}^{N-1}b_{1,j+1}\,z_{j+1} \left(-z_1+\frac{1}{2}[-b_{1,j+1}]_+z_{j+1}\right), 
$$ 
and
$$ 
G_L = -L(\upzeta )+\frac{1}{2}\log (1-\upzeta)\, \log\left(\frac{1-\upzeta}{\upzeta}\right) 
$$ 
for $\upzeta = \left(1+\exp(-  \sum_{k=1}^{N-1} b_{1,k+1}z_{k+1}) \right)^{-1}$. 
\end{propn} 
The proof of the preceding result follows from a direct calculation, using the 
conditions (\ref{reln1}) and (\ref{reln2}). (For another proof that $\varphi^* \om =\om$, 
see Lemma 2.3 in \cite{fh}.)

The two-form (\ref{omega}) is log-canonical: it is constant in the logarithmic coordinates $z_j$, 
so it is evidently closed, but may be degenerate. If det$\, B\neq 0$ then the map 
$\varphi$ is symplectic, but in general to obtain a symplectic map it is necessary to 
consider the null distribution of $\om$, which is generated by commuting vector fields 
of the form $\sum_j u_j x_j \frac{\partial}{\partial x_j}$, where the integer vector 
${\bf u}=(u_1,u_2,\ldots ,u_N)\in  \mathrm{ker}\, B$. Each such vector field integrates 
to yield a scaling transformation 
\beq\label{scale} 
{\bf x} \longrightarrow \la^{\bf u}\cdot {\bf x} = (\la^{u_1}x_1, \la^{u_2}x_2,\ldots, \la^{u_N}x_N), \qquad \la \in \C^*, 
\eeq 
so that overall there is an action of the algebraic torus $(\C^*)^{N-r}$, 
where $r=$rank$\, B$, 
and a complete set of independent invariants under these scaling transformations provides coordinates 
for the space of leaves of the null foliation for $\om$. 
Due to the skew-symmetry of the integer matrix $B$, $r$ is even, and the vector space 
$\Q^N$ has an orthogonal direct sum decomposition 
$\Q^N =  \mathrm{im}\, B \oplus \mathrm{ker}\, B$ with respect to the standard scalar product, denoted by $(\, , \,)$. 
Given an integer vector ${\bf v}=(v_1,\ldots ,v_N)$, the scaling action on the Laurent monomial ${\bf x}^{{\bf v}} = \prod_jx_j^{v_j}$
gives $\la^ {{\bf u}}\cdot {\bf x}^{{\bf v}} 
=\la^{({\bf u},{\bf v})}  {\bf x}^{{\bf v}}$, hence ${\bf x}^{{\bf v}}$ is 
invariant under the overall action of $(\C^*)^{N-r}$ if and only if $({\bf u},{\bf v})=0$ 
for all ${\bf u}\in \mathrm{ker}\, B$, so ${\bf v}\in\mathrm{im}\, B$. Thus a choice of basis 
$\{ {\bf v}_1, {\bf v}_2, \ldots , {\bf v}_r \}$ for $\mathrm{im}\, B$ defines a set of symplectic coordinates via the map 
\beq \label{pimap} 
\bear{lccl} 
\pi : \quad & \C^N & \longrightarrow &\C^r \\ 
              & {\bf x} & \longmapsto & {\bf U} := ({\bf x}^{{\bf v}_1}, {\bf x}^{{\bf v}_2}, \ldots, {\bf x}^{{\bf v}_r}) . 
\eear 
\eeq 
For what follows, it will be convenient to choose a $\Z$-basis for the 
$\Z$-module   $\mathrm{im}\, B_\Z=\mathrm{im}\, B\cap \Z^N$, which guarantees that $\varphi$ induces 
a birational map in the coordinates ${\bf U}=(U_1,U_2,\ldots ,U_r)$.  
(Different choices of basis are possible, such as taking any set of $r$ independent rows of $B$; cf. 
the $\tau$-coordinates in \cite{gsv, gsvduke}.)

\begin{thm} \label{torusred} 
Let $\{ {\bf v}_1, {\bf v}_2, \ldots , {\bf v}_r \}$ be 
a $\Z$-basis for 
$\mathrm{im}\, B_\Z$. Then given $\pi$ as in (\ref{pimap}), 
there is an associated symplectic
birational map
$\hat\varphi : \, \C^r\rightarrow \C^r$ such that 
$
\pi\cdot \varphi = \hat\varphi\cdot \pi , 
$ 
with 
$\pi^*\hat\om = \om$, 
where the 
symplectic form $\hat\om$ is log-canonical in the 
coordinates $(U_1,U_2,\ldots , U_r)$. 
\end{thm}
\begin{prf} 
This is equivalent to Theorem 2.6 in \cite{fh}, corresponding to a choice of 
basis as in case (a) of Lemma 2.9 therein.    
\end{prf} 

The rest of this section is taken up with presenting 
a technical result, Proposition~\ref{palilem} below, 
which shows that there is a special choice of $\Z$-basis in Theorem \ref{torusred} 
such that the map $\hat\varphi$ is equivalent to a recurrence relation, which 
we will refer to as the {\it U-system}. 

\begin{defn}\label{support}  
For a non-zero vector  $ {\bf v}\in\Q^N$, we say that $ {\bf v}$ has  {\em support} 
$\mathrm{supp}({\bf v})=[j,k]\subset [1,N]$ whenever $v_jv_k\neq 0$ and $v_i=0$ for all $i<j$ and $i>k$; 
and in that case the {\em length} of the support is $|\mathrm{supp}({\bf v})|=|[j,k]|=k-j+1$.  
Moreover, we say that a vector ${\bf v}$ with $\mathrm{supp}({\bf v})=[j,k]$ 
has {\em palindromic support} if $v_i = v_{k+j-i}$ for all $i\in [j,k]$. 
\end{defn} 

\bex \label{s4supp} 
{\em 
The exchange matrix 
(\ref{s4bmat}) has rows ${\bf b}_1, {\bf b}_2,{\bf b}_3,{\bf b}_4$ 
with $\mathrm{supp}({\bf b}_1)= [2,4]$, $\mathrm{supp}({\bf b}_2)= \mathrm{supp}({\bf b}_3)=[1,4]$ 
and $\mathrm{supp}({\bf b}_4)=[1,3]$. The vectors ${\bf b}_1$ and ${\bf b}_4$ 
have palindromic support, but  ${\bf b}_2$ and ${\bf b}_3$ do not.   
} 
\eex 

\begin{lem}\label{rsim} 
For  $ {\bf v}\in\Q^N$, define the reversal map $\mathrm{r}:\, {\bf v}=(v_j)\mapsto \mathrm{r}({\bf v})= (v_{N-j+1})$;  
and for $ {\bf v}$ with $\mathrm{supp}({\bf v})\subset [1,N-1]$ define the shift map 
$\mathrm{s}:\, {\bf v}=(v_j)\mapsto \mathrm{s}({\bf v})= (v'_{j})$,  where $v_1'=0$ and $v_j'=v_{j-1}$ for 
$j=2,\ldots ,N$. If 
${\bf v}\in  \mathrm{im}\, B$ then 
$\mathrm{r}({\bf v})\in  \mathrm{im}\, B$, 
and if also $\mathrm{supp}({\bf v})\subset [1,N-1]$ then 
$\mathrm{s}({\bf v})\in  \mathrm{im}\, B$ as well.  
\end{lem} 
\begin{prf}
Let  ${\bf b}_1, {\bf b}_2,\ldots,{\bf b}_N$ denote the rows of $B$. 
From the conditions (\ref{reln1}) and (\ref{reln2}) on the matrix elements 
of $B$, it follows  that $\mathrm{r}({\bf b}_j)=-{\bf b}_{N-j+1}$ for $j\in[1,N]$. 
Any ${\bf v}\in  \mathrm{im}\, B$ can be written as a
linear combination ${\bf v} = \sum_jc_j{\bf b}_j$, which means that 
the linear map r acts as $\mathrm{r}({\bf v})=-\sum_jc_j{\bf b}_{N-j+1}\in  \mathrm{im}\, B$, 
proving the first claim. 

For the second claim, note that whenever ${\bf v}\in  \mathrm{im}\, B$  with   $\mathrm{supp}({\bf v})\subset [1,N-1]$, 
the map $\varphi$ acts on the monomial ${\bf x}^{\bf v}$ to yield 
$\varphi^* (  {\bf x}^{\bf v}) = {\bf x}^{\mathrm{s}({\bf v})}$.  Picking a $\Z$-basis 
as in Theorem \ref{torusred} means that ${\bf v} = \sum_i w_i {\bf v}_i$ for a vector 
${\bf w}=(w_i)\in \Z^r$, and then 
$\varphi^* (  {\bf x}^{\bf v}) = \varphi^*\pi^* ({\bf U}^{\bf w})=\pi^*{\hat{\varphi}}^* ({\bf U}^{\bf w})$. 
So ${\bf x}^{\mathrm{s}({\bf v})}$ is the pullback of a function 
of the coordinates $U_i$,  implying  that this monomial is invariant 
under the action of the scaling transformations (\ref{scale}), and  
$({\bf u},\mathrm{s}({\bf v}))=0$, for all    ${\bf u}\in \mathrm{ker}\, B$. 
Hence $\mathrm{s}({\bf v})\in  \mathrm{im}\, B$, as required. 
\end{prf} 

\begin{defn} \label{palib}  
A {\em palindromic basis} for an $r$-dimensional subspace of $\Q^N$ is a 
basis $\{ {\bf v}_1, {\bf v}_2, \ldots , {\bf v}_r \}$ 
such that ${\bf v}_j =\mathrm{s}^{j-1}({\bf v})$, $j\in [1,r]$,  
for a vector ${\bf v}$ with palindromic support, where 
$\mathrm{supp}({\bf v})\subset [1,N-r+1]$.
\end{defn} 

\bex {\em 
For the matrix $B$ in Example \ref{s4bper}, the subspace 
$\mathrm{im}\, B\subset \Q^4$ has the palindromic 
basis $\{ {\bf v}_1, {\bf v}_2 \}$, where 
$
{\bf v}_1={\bf b}_4 = 
(1,-2,1,0)$ and 
${\bf v}_2= - {\bf b}_1 = 
(0,1,-2,1)$.
} 
\eex 

\begin{propn} \label{palilem} 
The subspace $\mathrm{im}\, B\subset \Q^N$ admits a palindromic basis, 
which yields a $\Z$-basis for $\mathrm{im}\, B_\Z$ (unique up to an 
overall sign). 
\end{propn} 
\begin{prf} See Appendix. \end{prf}

\bex \label{s6B} 
{ \em 
The $6\times 6$ exchange matrix 
\beq \label{s6bmat} 
B = \left(\bear{cccccc} 
0 & -1 & 1 & 0 & 1 & -1 \\ 
1 & 0 & -2 & 1 & -1 & 1 \\ 
-1 & 2 &  0 & -2 & 1 & 0 \\ 
0 & -1 & 2 & 0 & -2 & 1 \\ 
-1 & 1 & -1 & 2 & 0 & -1 \\ 
1 & -1 & 0 & -1 &  1 & 0 \eear \right) 
\eeq 
has ${\bf {a}} = (-1,1,0,1,-1)$, and the vectors 
${\bf {b}}_6 = (-{\bf {a}},0)$ and 
$\mathrm{s}({\bf {b}}_6) = (0,-{\bf {a}})=-{\bf {b}}_1$ span 
a 2-dimensional subspace of $\mathrm{im}\, B$, but 
$\mathrm{rank} \, B =4$. The vector 
${\bf v}=(1,-2,1,0,0,0)$ generates the 
palindromic $\Z$-basis 
$\{\, {\bf v}, \mathrm{s}({\bf v}),  \mathrm{s}^2({\bf v}), \mathrm{s}^3({\bf v})\,\}$ 
for $\mathrm{im}\, B_{\Z}$.
} 
\eex 

\begin{defn} \label{udef} 
Let $B$ be a skew-symmetric integer matrix satisfying the conditions 
(\ref{reln1}) and (\ref{reln2}), and let  
$\{ {\bf v}_1, {\bf v}_2, \ldots , {\bf v}_r \}$ be the unique 
palindromic $\Z$-basis  for $\mathrm{im}\, B_{\Z}$ such that 
the first component of ${\bf v}_1$ is positive. Then 
the iteration of the corresponding map $\hat\varphi$, 
as in Theorem \ref{torusred}, is equivalent to iterating a recurrence 
of the form 
\beq\label{usys} 
U_{n+r}\, U_n = {\cal F}(U_{n+1},U_{n+2},\ldots ,U_{n+r-1}),  
\eeq  
for a certain rational function ${\cal F}$. We refer to (\ref{usys}) 
as the {\em U-system} associated with the exchange matrix $B$. 
\end{defn} 

\bex\label{s4u} {\em 
The recurrence (\ref{s4qrt}), with $Y_n\to U_n$, is the U-system 
for the Somos-4 exchange matrix (\ref{s4bmat}). 
}
\eex 

\section{Y-systems and Z-systems} \label{ysys}

\setcounter{equation}{0}

In this section we give a rapid introduction to cluster algebras with coefficients, 
in order to describe Nakanishi's notion of periodicity 
in cluster algebras \cite{nakanishi}, which includes the 
cluster mutation-periodicity of \cite{fordy_marsh} as a special case. 
For all cluster algebras admitting such periodicity, it is 
possible to define associated T- and Y-systems. After a 
brief summary of the general situation, the focus returns to 
the case of cluster mutation-periodicity with period 1, as 
considered in the last section.

\subsection{Cluster algebras with coefficients} 

Let $\mathbb{P}$ be a semifield, an abelian multiplicative group 
endowed with a binary operation $\oplus$ which is
commutative, associative, and distributive 
with respect to the group multiplication. 
A {\it cluster algebra with coefficients}, of 
rank $N$, is 
the algebra $\mathcal{A}=\mathcal{A}(B,{\bf x},{\bf y})$ 
generated by the clusters of seeds 
$(B' ,{\bf x}',{\bf y}')$ produced from  
all possible sequences of mutations starting from an initial seed 
$(B ,{\bf x},{\bf y})$, where $B$ is an $N\times N$ skew-symmetrizable matrix, 
${\bf x}$ is a cluster, and ${\bf y}$ is a coefficient tuple 
${\bf y} = (y_1,\ldots,y_N) \in \mathbb{P}^N$.
%
%
Under the mutation $\mu_k$,  
the elements of $B$ are mutated according to (\ref{matmut}), 
while the coefficients have the exchange relation 
\beq \label{coex} 
\tilde{y}_j = 
\left\{ 
\begin{array}{ll}
y_k^{-1},\qquad & 
j=k, 
\\
y_j \Big(1\oplus y_k^{-\mathrm{sgn}(b_{kj}) }
\Big)^{-b_{kj}}, &   j\neq k. \end{array}
\right.  
\eeq 
The exchange relation for 
cluster variables is 
\beq \label{gexchange}
\tilde{x}_j = 
\left\{ 
\begin{array}{ll}
\displaystyle{\frac{y_k\prod_{j=1}^N x_j^{[b_{k,j}]_+} 
+ \prod_{j=1}^N x_j^{[-b_{k,j}]_+}}{(1\oplus y_k)x_k}},
\qquad & j = k,
\\[1mm]
x_i, & j \neq k,
\end{array}
\right.
\eeq
and this reduces to the relation (\ref{exch}) in the 
coefficient-free case by taking the projection from $\mathbb{P}$ 
to the trivial semifield consisting of a single element, $\{1\}$. 

Y-systems, to be introduced shortly, correspond to 
relations between elements of the universal semifield 
$\mathbb{P}_{univ}({\bf y})$, consisting of subtraction-free 
rational functions in the variables $y_j$. Working in this semifield, 
the addition $\oplus$ can be replaced by the usual addition $+$ 
in the field $\Q ({\bf y})$. 

\subsection{Periodicity and Y-systems} 

Nakanishi's general notion of periodicity can be stated as follows.
 
\begin{defn} 
Let the matrix $B$ belong to a seed in a cluster algebra $\mathcal{A}$ 
of rank $N$, and let ${\bf i}=(i_1,\ldots , i_h)$ be a sequence 
of indices $i_j\in I := [1,N]$. For the composition 
$\mu_{ {\bf i} }=\mu_{i_h}\cdot \ldots\cdot \mu_{i_2}\cdot\mu_{i_1}$, 
set $\tilde{B}=\mu_{ {\bf i} }B$, and let $\nu\in S_N$ be a 
permutation. Then the sequence  ${\bf i}$ is said to be a 
$\nu$-{\em period} of $B$ if $\tilde{B}=\nu(B)$, 
i.e. $\tilde{b}_{\nu(j)\nu(k)}=b_{jk}$ for 
$j,k=1,\dots,N$; moreover, if $\nu =${\em id} then it is just called 
a {\em period} of $B$. 
\end{defn} 

\bex {\em 
A cluster mutation-periodic quiver $Q$ with period 
$m$, as defined in  \cite{fordy_marsh}, is equivalent 
to a skew-symmetric matrix $B$ for which the sequence 
${\bf i} = (1,2,\ldots,m)$ is $\rho^m$-periodic,  for 
a cyclic permutation $\rho$, 
in the sense of the above definition.  
In particular, a period 1 quiver 
is equivalent to a matrix $B$ for which 
${\bf i}=(1)$ is a $\rho$-period; any such matrix satisfies the 
conditions (\ref{reln1}) and (\ref{reln2}). 
} 
\eex    

\begin{rem} {\em 
Note that \cite{fordy_marsh} and \cite{nakanishi} 
adopt opposite conventions for labelling permutations, 
corresponding to $\rho\leftrightarrow \rho^{-1}$ 
above. 
In this section we follow \cite{nakanishi}. 
} 
\end{rem} 
 
For a symmetrizable matrix $B$ with a $\nu$-{\em period} ${\bf i}$, 
the Y-system is, roughly speaking, a set of algebraic relations between the 
coefficients $\tilde{y}_i$ obtained by applying mutations
$\mu_{\bf i}, ~\mu_{\nu({\bf i})}, ~\ldots,~ \mu_{\nu^{g-1}({\bf i})}$
in this order, where $g$ is the order of $\nu$.  
To be more precise, we recall some definitions in \cite{nakanishi}.
\begin{defn}
Let $B$ be a symmetrizable matrix with a $\nu$-{\em period} ${\bf i}$.
We say ${\bf i}$ is {\em regular} if 
all the components of ${\bf i}|\nu({\bf i})|\cdots|\nu^{g-1}({\bf i})$
exhaust $I$, and if all the components of ${\bf i}$ belong to distinct
$\nu$-orbits in $I$.
We decompose ${\bf i}$ into $t$ parts,
${\bf i}(0) | {\bf i}(1) | \cdots | {\bf i}(t-1)$
and write ${\bf i}(p)=(i(p)_1,\ldots,i(p)_{r_p})$ for $p=0,\ldots,t-1$.
This decomposition is called a {\em slice} of ${\bf i}$, of length $t$,
if the mutations $\mu_{i(p)_1},\ldots,\mu_{i(p)_{r_p}}$
form a commuting set for each $p$.
\end{defn}

We assume that ${\bf i}$ is regular and has a slice
${\bf i}={\bf i}(0)| {\bf i}(1)|\ldots |{\bf i}(t-1)$.
Then we define a sequence of seeds 
$\{(B(u),{\bf x}(u), {\bf y}(u))\}_{u \in \mathbb{Z}}$ by
\begin{eqnarray*}
  \cdots & \cdots
  \stackrel{\mu_{\nu^{-1}({\bf i}(t-2))}}{\longleftrightarrow}
  (B(-1),{\bf x}(-1), {\bf y}(-1)) 
  \stackrel{\mu_{\nu^{-1}({\bf i}(t-1))}}{\longleftrightarrow}
  \\
  &(B(0),{\bf x}(0), {\bf y}(0)) 
  \stackrel{\mu_{{\bf i}(0)}}{\longleftrightarrow}
  (B(1),{\bf x}(1), {\bf y}(1))
  \stackrel{\mu_{{\bf i}(1)}}{\longleftrightarrow}
  \cdots
  \stackrel{\mu_{{\bf i}(t-1)}}{\longleftrightarrow}
  \\
  &(B(t),{\bf x}(t), {\bf y}(t)) 
  \stackrel{\mu_{\nu({\bf i}(0))}}{\longleftrightarrow}
  (B(2),{\bf x}(t+1), {\bf y}(t+1))
  \stackrel{\mu_{\nu({\bf i}(1))}}{\longleftrightarrow}
  \cdots \cdots
\end{eqnarray*}
where $(B(0),{\bf x}(0), {\bf y}(0)) := (B,{\bf x}, {\bf y})$. 
The set of {\em forward mutation points} $P_+\subset I \times \Z$ 
are pairs $(i,u)\in I\times \Z$ such that $i$ is a component of $\nu ^m ({\bf i}(k))$ for $u=mt +k$ ($m\in\Z$, 
$k\in [0,t-1]$). Take $g_i$ to be the smallest positive integer such 
that $\nu^{g_i}(i)=i$. Then the Y-system takes the form 
%
%
\beq \label{ysysi} 
y_i(u)\,y_i(u+tg_i)=\frac{\prod_{(j,v)\in P_+} (1+y_j(v))^{G'_+(j,v;i,u)} } 
{\prod_{(j,v)\in P_+} (1+y_j(v)^{-1})^{G'_-(j,v;i,u)} },   
\eeq 
where 
$$
G_{\pm}'(j,v;i,u) = 
\left\{
\begin{array}{ll}
\mp b_{ji}(v) \qquad & \mathrm{if} \quad
v\in (u,u+tg_i), \,  b_{ji}(u) \lessgtr 0, \\
0 & \mathrm{otherwise} . \end{array}
\right.
$$ 
The corresponding (coefficient-free) T-system is 
\beq \label{tsysi} 
x_i(u)\,x_i(u+tg_i)=\prod_{(j,v)\in P_+} x_j(v)^{H'_+(j,v;i,u)} + 
\prod_{(j,v)\in P_+} x_j(v)^{H'_-(j,v;i,u)} , 
\eeq 
where 
$$
H_{\pm}'(j,v;i,u) = 
\left\{
\begin{array}{ll}
\pm b_{ji}(u) \qquad & \mathrm{if} \quad
u\in (v-tg_j,v), \,  b_{ji}(u) \gtrless0, \\
0 & \mathrm{otherwise} . \end{array}
\right.
$$ 

Here we modify the T-system by introducing a coefficient that multiplies both terms on the right 
hand side. (However, this is different from the way that coefficients appear 
in the general exchange relation (\ref{gexchange}) in a cluster algebra.)

\begin{propn} \label{tz} 
Let $x_i(u)$ satisfy the modified T-system 
\begin{eqnarray} 
\label{mtsysi} 
\fl \quad x_i(u)\,x_i(u+tg_i)={\cal Z}_i(u)\Big(\prod_{(j,v)\in P_+} x_j(v)^{H'_+(j,v;i,u)} + 
\prod_{(j,v)\in P_+} x_j(v)^{H'_-(j,v;i,u)} \Big).
\end{eqnarray} 
Then 
\beq\label{tysub} 
\overline{y}_i(u)=  \prod_{(j,v)\in P_+} \frac{x_j(v)^{H'_+(j,v;i,u)} }{x_j(v)^{H'_-(j,v;i,u)}}
\eeq 
satisfies the Y-system (\ref{ysysi}) if and only if 
\beq\label{zsysi} 
\prod_{(j,v)\in P_+} \frac{{\cal Z}_j(v)^{H'_+(j,v;i,u)} }{{\cal Z}_j(v)^{H'_-(j,v;i,u)}}=1. 
\eeq 
\end{propn} 
\begin{prf} Upon substituting $y_i(u)=\overline{y}_i(u)$ into 
each side of (\ref{ysysi}), this follows from a direct calculation that is 
almost identical to the proof of Proposition 5.11 in \cite{nakanishi}.   
\end{prf} 

We refer to (\ref{zsysi}) as the {\em Z-system}, while we say that the Z-system together 
with the modified T-system (\ref{mtsysi}) in Proposition \ref{tz} constitute the {\em T$_z$-system}.   

Henceforth we focus on the special case of cluster mutation-periodicity with period 1, which we 
describe in more detail. 
In order to define the Y-system in the period 1 case, one starts from $B(0)=B$ 
and defines a sequence $B(u)$ for $u\in\Z$ by applying an infinite sequence of mutations, 
starting with $B(1)=\mu_{1}(B(0))$, 
then  $B(2)=\mu_{\rho (1)}(B(1))$, 
$B(3)=\mu_{\rho^2 (1)}(B(2))$, and so on, and similarly 
extending backwards with $B(-1)=\mu_{\rho^{-1}(1)}(B(0))$, etc. 
In this way, one has $B(n)=\rho^n(B)$ for $n\in\Z$, 
hence $B(N)=B$ since $\rho^N=$id; so this sequence of matrices is periodic. 
Similarly, this infinite sequence of mutations produces a sequence 
of clusters ${\bf x}(u)$  and a sequence of coefficient tuples 
${\bf y}(u)$ for 
$u\in\Z$, but in general the latter two sequences are not periodic.

 In the period 1 case, all the vertices of the quiver 
corresponding to $B$ cycle with period $N$ under the action of $\rho$, 
and the Y-system relates $y_j(u)$ to $y_j(u+N)$. 
In fact, in this case, by replacing 
$y_j(n-1)\to y_n$ 
whenever $n=j \, \bmod \, N$, 
one can write a single recurrence 
relation for $y_n$, $n\in\Z$, as follows.         
 
\begin{propn} 
The Y-system associated with a cluster mutation-periodic quiver $Q$ 
with period 1 corresponding to a skew-symmetric matrix $B$ can be written 
as the recurrence  
\beq\label{ysystem} 
{y}_{n+N}\,{y}_n = 
\frac{\prod_{j=1}^{N-1}(1+{y}_{n+j})^{[-b_{1,j+1}]_+}} 
{\prod_{j=1}^{N-1}(1+{y}_{n+j}^{-1})^{[b_{1,j+1}]_+}}.  
\eeq 
\end{propn} 
\begin{prf} This 
follows from (\ref{ysysi}), which is the same as equation (5.8) in \cite{nakanishi}. 
\end{prf} 

Note that (\ref{ysystem}) is the same as (\ref{ayrec}), with the exponents written in terms of the 
coefficients in the first row of $B$. 

The T-system with coefficients corresponding to the $\rho$-period $(1)$  
consists of the analogous set of formulae 
for the combinations $x_j(u+N)x_j(u)$ of cluster variables, 
generated by the exchange relation (\ref{gexchange}), and the coefficient-free case 
is induced by the projection 
$\pi_1: \, \mathbb{P}_{univ}({\bf y})\rightarrow \{1\}$ from the coefficient semifield 
to the trivial semifield. 
As for $y_j(u)$, the period 1 property of $B$ means that we can replace $x_j(n-1)\to x_n$ 
whenever $n=j \, \bmod \, N$, and write a single recurrence for $x_n$. 

\begin{propn}\label{tsubs} 
In the case of cluster mutation-periodicity with period 1, 
the T-system without coefficients 
is equivalent to the recurrence (\ref{crec}).  
If  $x_n$ is a solution of (\ref{crec}), 
then taking $y_n=\bar{y}_n$, where  
\beq \label{tsubst} 
\bar{y}_n =\prod_{j=1}^{N-1} {x}_{n+j}^{-b_{1,j+1}} ,   
\eeq 
yields a solution of the Y-system (\ref{ysystem}).   
\end{propn} 
\begin{prf}
This is a special case of 
Proposition 5.11 in \cite{nakanishi}, using the formula (5.29) therein.   
\end{prf}

As we saw in section 2, for the example of Somos-4,  the substitution (\ref{tsubst}) does not 
provide the general solution of the Y-system (\ref{ysystem}). The key to understanding 
the discrepancy between the T-system and the Y-system is the following. 

\begin{propn}\label{tzsubs} 
Let $x_n$ be a solution of the modified T-system 
\beq\label{czrec}
x_{n+N}x_n = {\cal Z}_n\left(\prod_{j=1}^{N-1} x_{n+j}^{[b_{1,j+1}]_+} + \prod_{j=1}^{N-1} x_{n+j}^{[-b_{1,j+1}]_+}\right).
\eeq 
Then taking 
$y_n=\bar{y}_n$, where  
\beq \label{tzsub} 
\bar{y}_n =\prod_{j=1}^{N-1} {x}_{n+j}^{-b_{1,j+1}} ,   
\eeq 
yields a solution of the Y-system if and only if ${\cal Z}_n$ satisfies 
\beq \label{zsys} 
\prod_{j=1}^{N-1} {\cal Z}_{n+j}^{-b_{1,j+1}}=1 .    
\eeq 
The space of solutions of the T$_z$-system consisting of (\ref{czrec}) together with (\ref{zsys}) 
has dimension $N+\tilde{r}$, where $\tilde{r}=|\mathrm{supp}({\bf b}_1)|-1$. 
\end{propn}   
\begin{prf} The 
first claim above is just a special case of Proposition  \ref{tz}. To prove the second claim, observe 
that the exponents appearing in (\ref{zsys}) are (up to sign) just the components of the vector 
${\bf b}_1$ in the first row of $B$, so that in this case the Z-system is a difference equation 
of order $\tilde{r}$. Then $\tilde{r}$ initial values are required for (\ref{zsys}), say 
${\cal Z}_0, {\cal Z}_1,\ldots , {\cal Z}_{\tilde{r}-1}$, together 
with a further $N$ values $x_0,x_1,\ldots,x_{N-1}$ in (\ref{czrec}), giving a total of $N+\tilde{r}$ initial data.
\end{prf}
\begin{rem}{\em 
In general, unless the first non-zero component of ${\bf b}_1$ 
(the leading exponent) is $\pm 1$, the Z-system (\ref{zsys}) 
is an algebraic recurrence relation rather than a rational one. When the leading exponent is 
$\pm 1$, the Z-system just generates a sequence of monomials in the initial data, 
and in all such examples it appears that the Laurent property holds for 
(\ref{czrec}), in the sense that 
$x_n \in \Z [x_0^{\pm 1}, \ldots ,  x_{N-1}^{\pm 1}, {\cal Z}_0^{\pm 1}, \ldots , {\cal Z}_{\tilde{r}-1}^{\pm 1}]$.  
}
\end{rem} 

The discrepancy between the solutions of the Y-system and the coefficient-free T-system 
can be seen by looking at the fibres of the map from ${\bf x}$ variables 
to ${\bf y}$ variables defined by (\ref{tzsub}): generically, these have 
dimension $\tilde{r}$, as there is an action of 
a torus $(\C^*)^{\tilde{r}}$ by scalings ${\bf x} \to \la^{\bf u}\cdot {\bf x}$ 
for integer vectors ${\bf u}$ that are orthogonal to ${\bf b}_1$ and  
the shifts s$^k({\bf b}_1)$ (for all $k$ where this is defined), and the value 
of ${\bf y}$ is preserved by any such scaling. The solutions 
of the Y-system can also be understood by introducing symplectic coordinates 
in the same way as for the coefficient-free T-system, which leads to the 
following. 

\begin{defn} \label{uzdef} 
The {\em U$_z$-system} associated with (\ref{czrec}) consists of the recurrence 
\beq\label{uzsys} 
U_{n+r}\, U_n = {\cal Z}_n \, {\cal F}(U_{n+1},U_{n+2},\ldots ,U_{n+r-1}),  
\eeq  
with the same rational function ${\cal F}$ as in (\ref{usys}), together with (\ref{zsys}). 
\end{defn}

\section{Examples of T$_z$-systems and equations of q-Painlev\'e type} 
\setcounter{equation}{0}

In this section we give some examples of T$_z$-systems and U$_z$-systems 
for some particular Y-systems associated with quivers that are mutation periodic with 
period 1. 

\subsection{Affine A-type quivers and linear relations} 

The $4\times 4$ exchange matrix 
$$ 
B = \left(\bear{cccc} 
0 & -1 & 0 & -1 \\ 
1 & 0 & -1 & 0 \\ 
0 & 1  & 0 & -1 \\ 
1 & 0 &  1 & 0 \eear \right) 
$$ 
corresponds to the $A_3^{(1)}$ affine Dynkin diagram, oriented such that 
one of the  arrows is clockwise and the other three are anticlockwise;  
this quiver is sometimes denoted $\tilde{A}_{1,3}$, and it is also 
one of the primitive quivers involved in the classification of 
period 1 quivers by Fordy and Marsh \cite{fordy_marsh}, who denote it $P_4^{(1)}$.    
The T$_z$-system in this case has the form 
\beq \label{prim4t} 
x_{n+4}\, x_n = {\cal Z}_n \, (x_{n+3}\, x_{n+1} + 1), \qquad \mathrm{with} \qquad {\cal Z}_{n+2}\, {\cal Z}_n =1. 
\eeq 
The simple form of the Z-system above means that the coefficients of the non-autonomous T-system 
are periodic, being 
given by the sequence ${\cal Z}_0,{\cal Z}_1, {\cal Z}_0^{-1},{\cal Z}_1^{-1}$, 
repeated with period 4. Since the matrix $B$ is of full rank, the  U$_z$-system 
is the same as the T$_z$-system in this case, while the Y-system is given by 
\beq \label{prim4y} 
y_{n+4}\, y_n = (1+y_{n+3})(1+y_{n+1}). 
\eeq 

In \cite{honelaur} it was shown that for the autonomous version of 
the T-system, 
obtained by taking ${\cal Z}_n=1$ for all $n$ in (\ref{prim4t}), the iterates 
satisfy the fourth-order linear recurrence 
$x_{n+4} - K x_{n+2} + x_n =0$,  where the coefficient $K$ is 
a first integral (conserved quantity), constant along orbits. 
Using computer algebra (with Maple) we have verified the following 
result for the   T$_z$-system. 
\begin{propn} \label{prim4lin} 
The iterates $x_n$ of the T$_z$-system (\ref{prim4t}) 
satisfy the linear relation 
\beq 
\label{prim4rel}
x_{n+24}-\mathrm{C}\, x_{n+12} + x_n =0, 
\eeq 
where the first integral 
$
\mathrm{C}
$ 
is a Laurent polynomial in $\Z [x_0^{\pm 1}, x_1^{\pm 1},x_2^{\pm 1},x_3^{\pm 1},{\cal Z}_0^{\pm 1},{\cal Z}_1^{\pm 1}]$ 
consisting of 67 terms with positive coefficients. 
\end{propn} 
\noindent 
The existence of such a linear relation means that 
(\ref{prim4t}), and hence (\ref{prim4y}), can be solved 
explicitly, and from the form of (\ref{prim4rel}) the solution can be 
written in terms of Chebyshev polynomials with argument $\mathrm{C}/2$.  

For cluster algebras obtained from affine Dynkin quivers,  
linear relations between cluster variables have been derived   
in various ways \cite{assem,fordy_marsh,fordy_rec,keller_scher}.
In particular one can take the $A_{N-1}^{(1)}$ Dynkin diagram with 
one clockwise arrow and $N-1$ anticlockwise arrows to 
get the quiver  $\tilde{A}_{1,N-1}$, which is the same 
as the  primitive $P_N^{(1)}$ in  \cite{fordy_marsh}. 
For this family of quivers,  
the T$_z$-system has the form 
\beq \label{primNt} 
x_{n+N}\, x_n = {\cal Z}_n \, (x_{n+N-1}\, x_{n+1} + 1), \qquad \mathrm{with} \qquad {\cal Z}_{n+N-2}\, {\cal Z}_n =1, 
\eeq 
while the Y-system is 
\beq \label{primNy} 
y_{n+N}\, y_n = (1+y_{n+N-1})(1+y_{n+1}). 
\eeq 
Extensive numerical experiments suggest that there 
are linear relations with constant coefficients for 
all members of this family, given as follows. 
\begin{conje}\label{linaconj} 
For each $N\geq 3$, the iterates of the $\tilde{A}_{1,N-1}$ T$_z$-system (\ref{primNt}), associated 
with the Y-system  (\ref{primNy}), 
satisfy a constant-coefficient linear recurrence relation of order $4(N-1)(N-2)$, 
having the form 
$$ 
x_{n+4s}- \mathrm{A}\,  x_{n+3s} + \mathrm{B} \, x_{n+2s} -\mathrm{A} \, x_{n+s} +x_n =0  
$$ 
for $N$ odd, and 
$$ 
x_{n+4s}- \mathrm{C}\, x_{n+2s} + x_n =0  
$$   
for $N$ even, 
where $\mathrm{A,B,C}$ denote first integrals, and $s=(N-1)(N-2)$.  
\end{conje}

\subsection{Somos-type T$_z$-systems} 

Rather than considering the most general Somos-type recurrence, of 
the form (\ref{somosN}), here we  focus on some particular examples 
which indicate how the results in section 2 generalize to 
other q-Painlev\'e type equations and their higher-order analogues. 

To begin with, we look at the family of  bilinear 
T$_z$-systems given by 
\bea\label{somNtz} 
\fl \quad x_{n+N}\, x_n = {\cal Z}_n \, (x_{n+N-1}\, x_{n+1} + x_{n+N-2}\, x_{n+2}), 
\qquad \frac{{\cal Z}_{n+N-2}\, {\cal Z}_n}
{ {\cal Z}_{n+N-3}\, {\cal Z}_{n+1}}=1, 
\eea 
corresponding to $p=1$, $q=2$ in (\ref{somosN}). After solving the Z-system (\ref{doll}), the case $N=4$ 
of (\ref{somNtz}) is equivalent to 
(\ref{bildpi}). In general, just as for Somos-4,   
it is always possible to make a gauge transformation 
to move the non-autonomous coefficient entirely onto either the first or the second term on the right-hand side 
of the recurrence for $x_n$; we illustrate this in some specific examples below. 
A special case of the family (\ref{somNtz}), with purely periodic coefficients, 
arose by a 
reduction of the discrete KdV equation in \cite{dkdv}.  
\bex [Somos-5] 
{\em
\label{s5}
For 
$N=5$, 
the T$_z$-system 
(\ref{somNtz}) becomes 
$$ 
x_{n+5}\, x_n = {\cal Z}_n \, (x_{n+4}\, x_{n+1} + x_{n+3}\, x_{n+2}), 
\qquad \frac{{\cal Z}_{n+3}\, {\cal Z}_n}
{ {\cal Z}_{n+2}\, {\cal Z}_{n+1}}=1
,
$$ 
which corresponds to the Y-system 
$$ 
y_{n+5}\, y_n  = 
\frac{(1+{y}_{n+4}) (1+{y}_{n+1})} 
{(1+{y}_{n+3}^{-1}) (1+{y}_{n+2}^{-1})}.  
$$ 
The exchange matrix $B$ in this case has rank 2 (see \cite{sigma} for details), and taking $U_n = x_{n+3}x_n/(x_{n+2}x_{n+1})$ 
gives the associated U$_z$-system 
\beq \label{s5u} 
U_{n+2}\, U_n = 
{\cal Z}_n \, 
(1+U_{n+1}^{-1}), 
\qquad 
\mathrm{with} 
\qquad 
{\cal Z}_{n} = \beta_n \, \mathfrak{q}^n, \quad \beta_{n+2} = \beta_n. 
\eeq 
Each iteration of the latter is symplectic, preserving the log-canonical 
2-form $\hat\om = ({U}_n \, {U}_{n+1})^{-1} \, \dd {U}_{n+1} \wedge \dd {U}_{n}$, and 
according to \cite{asymmetric_dp} (see  (24) therein) the equation 
(\ref{s5u}) is a version of q-Painlev\'e II. 
Note that it is possible to use a gauge transformation $G_n$ in 
the T$_z$-system, setting $x_n = G_n x'_n$, to move all of the non-autonomous part onto the first 
term on the right-hand side; in that case, by taking $U'_n =  x'_{n+3}x'_n/(x'_{n+2}x'_{n+1})$, 
the equation (\ref{s5u}) can be transformed to 
$$ 
U'_{n+2}\, U'_n =  \al_n +\beta \, (U_{n+1}')^{-1}, \qquad \mathrm{where} \qquad 
({\cal S}^3 -1) ({\cal S}^2 -1)({\cal S} +1)\log\al_n = 0  \, \bmod\, 2\uppi \mathrm{i},  
$$ 
with ${\cal S}$ denoting the shift operator and the coefficient $\beta$ being constant. 
(The autonomous case $\al_n=\,$constant is a QRT map, explicitly solved in \cite{hones5}.) 
Observe that this choice of gauge introduces period 3 behaviour which is not present 
in  ${\cal Z}_{n}$. 
} 
\eex 
\bex [A particular case of Somos-6] 
{\em
\label{s6}
For 
$N=6$, 
the T$_z$-system 
(\ref{somNtz}) is 
$$ 
x_{n+6}\, x_n = {\cal Z}_n \, (x_{n+5}\, x_{n+1} + x_{n+4}\, x_{n+2}), 
\qquad \frac{{\cal Z}_{n+4}\, {\cal Z}_n}
{ {\cal Z}_{n+3}\, {\cal Z}_{n+1}}=1
,
$$ 
which corresponds to the exchange matrix (\ref{s6bmat}), and gives the Y-system 
$$ 
y_{n+6}\, y_n  = 
\frac{(1+{y}_{n+5}) (1+{y}_{n+1})} 
{(1+{y}_{n+4}^{-1}) (1+{y}_{n+2}^{-1})}.  
$$ 
From Example \ref{s6B}, the matrix $B$ has rank 4, and setting 
$U_n =x_nx_{n+2}/(x_{n+1})^2$ in the T$_z$-system and solving the 
Z-system produces 
the 
U$_z$-system 
$$ 
U_{n+4}(U_{n+3} U_{n+2} U_{n+1})^2 U_n =  
\beta_n \, \mathfrak{q}^n \, 
(1+U_{n+3} U_{n+2} U_{n+1}), 
\quad 
\mathrm{with} 
\quad 
 \beta_{n+3} = \beta_n. 
$$ 
Each iteration of the latter is a symplectic map, preserving the nondegenerate Poisson 
bracket in four dimensions defined by 
$$ 
\{ U_n,U_{n+1}\}= U_nU_{n+1}, \quad 
\{ U_n,U_{n+2}\}= -U_nU_{n+2}, \quad 
\{ U_n,U_{n+1}\}= U_nU_{n+3}.  
$$ 
In the autonomous case ($\mathfrak{q}=1$ and $\beta_n=\,$constant) 
this is equivalent to one of the Liouville integrable maps in \cite{hones6}.  
}
\eex

The preceding two examples can be generalized to cover this whole 
family, depending on whether $N$ is even/odd. 
For each $N$, the solution of the Z-system in (\ref{somNtz}) can be written as 
$$ 
{\cal Z}_{n} = \beta_n \, \mathfrak{q}^n, \qquad \beta_{n+N-3} = \beta_n;  
$$ 
when $N$ is even, the substitution $U_n =x_nx_{n+2}/(x_{n+1})^2$ gives 
the 
U$_z$-system 
\beq\label{sNeven} 
U_{n+N-2}\left(\prod_{j=1}^{N-3} U_{n+j}\right)^2 U_n =  
 {\cal Z}_{n}  \, 
\left(1+ \prod_{k=1}^{N-3} U_{n+k} \right), 
\eeq 
while for $N$ odd, setting $U_n =x_nx_{n+3}/(x_{n+1}x_{n+2})$ yields 
the 
U$_z$-system 
\beq\label{sNodd} 
\prod_{j=0}^{N-3} U_{n+j} =  
 {\cal Z}_{n}  \, 
\left(1+ \prod_{k=0}^{(N-5)/2} U_{n+2k+1} \right). 
\eeq 

We consider one further Somos-type example, which is outside the family 
(\ref{somNtz}). 

\bex [A special case of Somos-7] {\em

\label{som7}
One of the examples considered in section 6 of \cite{fh} 
corresponds to the T-system for a $7\times 7$ exchange matrix  
with rank $B=2$, reducing to a symplectic map of the plane. 
The associated T$_z$-system  is 
\beq
\label{somos7}
x_{n+7}\, x_n = {\cal Z}_{n}(x_{n+6}\, x_{n+1}+ x_{n+4}\, x_{n+3}), \qquad 
\frac{{\cal Z}_{n+5}\, {\cal Z}_n}
{ {\cal Z}_{n+3}\, {\cal Z}_{n+2}}=1
, 
\eeq
with the Y-system being 
$$
y_{n+7}\, y_n = \frac{(1+{y}_{n+6}) (1+{y}_{n+1})} 
{(1+{y}_{n+4}^{-1}) (1+{y}_{n+3}^{-1})} 
,
$$ 
while setting $U_n = x_{n+5}x_n/(x_{n+3}x_{n+2})$ yields 
the U$_z$-system 
$$ 
U_{n+2}\, U_n =   {\cal Z}_{n}\, (U_{n+1} + 1). 
$$ 
In this case, it is convenient to choose a gauge transformation $G_n$, with $x_n = G_n x'_n$, which 
moves all of the non-autonomous part onto the coefficient of the second term  on the right-hand side 
(and the coefficient of the first term can be fixed to be 1). This gives an equation 
for     $U'_n = x'_{n+5}x'_n/(x'_{n+3}x'_{n+2})$ which is of Lyness type, that is 
\bea\label{lynessn} 
\fl \quad U'_{n+2}\, U'_n =   U'_{n+1} + a_n,   
\quad \mathrm{with} \quad 
({\cal S} -1)({\cal S}^6 -1) \log a_n = 0  \, \bmod\, 2\uppi \mathrm{i} . 
\eea 
The form of the above is consistent with the results in 
\cite{cima}, which show that, in the case that $a_n$ is periodic,  
the Lyness recurrence (\ref{lynessn}) is integrable when the period 
is a divisor of 6, and indicate that chaos is present for all 
other periods. However,  the 
characteristic polynomial of the linear recurrence for $\log a_n$ 
has 1 as a double root, so in general  
$a_n =\al_n \, \mathfrak{q}^n$ with $\al_n$ being periodic with period 6, 
and (\ref{lynessn}) is yet another q-Painlev\'e equation: 
it is equivalent to equation (28) in \cite{asymmetric_dp}, where it is identified as a 
Schlesinger transformation for (20), 
a form of q-Painlev\'e V.  
} 
\eex 

\subsection{A nonintegrable example} 

The examples considered so far in this section should all be integrable 
in some sense. In the context of iteration of rational functions, one way to characterize integrability of discrete dynamical 
systems is in terms of the weak growth of degrees of iterates: defining 
the algebraic entropy as ${\cal E}:=\lim_{n\to\infty} n^{-1} \log d_n$, where $d_n$ is the degree of 
the $n$th iterate,  discrete integrable systems 
should have zero entropy  \cite{bellon_viallet}. In \cite{fh}, one of us used 
the growth of denominators of ${\bf x}$ variables to identify 
integrable maps among the T-systems (\ref{arec}); this growth is 
determined by the tropical version of the T-system (in the sense of the 
max-plus algebra).  All members of the families  
(\ref{sNeven}) and (\ref{sNodd}) are expected to have zero entropy, 
since the associated T$_z$-systems (\ref{somNtz}) have tropical versions that are almost identical 
to those of their autonomous forms, discussed in section 6 of \cite{fh}.   
Yet integrability is a rare phenomenon: for the majority of the T-systems (\ref{arec}), 
unless the exponents $a_j$ are of sufficiently small magnitude, the degrees 
grow exponentially with $n$, indicating nonintegrability. Likewise, the 
corresponding       T$_z$-system also exhibits exponential growth of degrees  
in the generic case, and usually this can be detected by looking at the 
Z-system alone, which we illustrate here with one particular example.  

For the exchange matrix  
\beq\label{6b} %
B=\left(
\begin{array}{cccccc} 0 & -2 & 6 & -4 & 6 & -2 \\
                      2 & 0 &  -14 & 6 & -16 & 6 \\
                     -6 & 14 & 0  &  10 & 6 & -4 \\
                      4 & -6 & -10 & 0 & -14 & 6 \\
                      -6 & 16 & -6 & 14 & 0 & -2 \\
                       2 & -6 & 4 & -6 & 2 & 0 \eear\right), 
\eeq 
%
%
the T$_z$-system  is 
\beq
\label{sixthordernew}
x_{n+6}\, x_n = {\cal Z}_{n}\, (x_{n+5}^2 x_{n+3}^4 x_{n+1}^2+ (x_{n+4}\, x_{n+2})^6).
\eeq
with 
\beq\label{6nz} 
\frac{{\cal Z}_{n+4}^2 \, {\cal Z}_{n+2}^4 \,{\cal Z}_n^2}
{ {\cal Z}_{n+3}^6\, {\cal Z}_{n+1}^6}=1 , 
\eeq 
and the associated Y-system is 
$$ 
y_{n+6}\, y_n =  \frac{(1+{y}_{n+5})^2 (1+{y}_{n+1})^4 (1+{y}_{n+1})^2} 
{(1+{y}_{n+4}^{-1})^6 (1+{y}_{n+2}^{-1})^6}. 
$$ 
The autonomous version of (\ref{sixthordernew}) was 
presented as Example 2.6 in \cite{sigma}: the matrix (\ref{6b}) has rank 2, 
with 
${\bf v}= (1,-3,2-3,1,0)$ and  $\mathrm{s}({\bf v})=  (0,1,-3,2-3,1)$ 
providing a palindromic 
$\Z$-basis for $\mathrm{im}\, B_\Z$, which leads 
to the U$_z$-system 
$$ 
U_{n+2}\, U_n =   {\cal Z}_{n}\, (U_{n+1}^{-1} + U_{n+1}^{-3}),  
$$ 
where ${\cal Z}_n$ satisfies (\ref{6nz}). 

The first thing to note about this example is that the Z-system 
involves indeterminacy due to the leading exponent 2: 
to iterate (\ref{6nz}) requires taking a square root to 
solve for ${\cal Z}_{n+4}$, so that a $\pm$ sign must be chosen 
coherently at each step. The second main observation is that, up to sign choices, the explicit solution to the Z-system 
can be obtained by solving the linear recurrence   
\beq \label{6nlin} 
({\cal S}^2 +1)({\cal S}^2 -3{\cal S}+1) \log {\cal Z}_{n} = 0  \, \bmod\, \uppi \mathrm{i} .
\eeq 
The nonintegrable nature of the system can be seen from the second quadratic factor 
above, which has a characteristic root $\la_{max} = (3+\sqrt{5})/2$, of magnitude 
greater than 1; this is 
in accordance with the positive value 
${\cal E}= \log \la_{max} $ for the entropy of the corresponding T-system, as 
found in Example 3.2 of \cite{sigma}. Taking ${\cal Z}_{0},{\cal Z}_{1},{\cal Z}_{2},{\cal Z}_{3}$
as initial data, the  Z-system   
(\ref{6nz}) generates monomials of the form 
$$ 
{\cal Z}_{n}= \pm \, \frac{{\cal Z}_{3}^{d_n^{(3)}}{\cal Z}_{1}^{d_n^{(1)}}}{{\cal Z}_{2}^{d_n^{(2)}}{\cal Z}_{0}^{d_n^{(0)}}}, 
$$ 
where each exponent $d_n^{(j)}$ for $j=0,1,2,3$ satisfies the same homogeneous linear recurrence 
that is defined by the difference operator in (\ref{6nlin}); hence the degrees 
of these monomials grow like a constant times $\la_{max}^n$. 

Our experience with other T$_z$-systems obtained from cluster mutation-periodic 
quivers with periods 1 and  2 suggests that if the Z-system displays exponential growth, 
and/or it has a leading exponent greater than 1, then the underlying T-system 
should 
have positive entropy. However, in general (unlike the preceding example) 
the degrees of ${\bf x}$ variables and  ${\cal Z}$ variables need not 
grow with the same rate. 

\section{Conclusions} 

We have shown that there is a link between periodicity in cluster algebras, Y-systems and  discrete Painlev\'e equations. 
This link also produces  higher-order analogues of q-Painlev\'e equations, such as 
the families (\ref{sNeven}) and (\ref{sNodd}), which to the best of our knowledge are new.   
That there is such a link should not be entirely surprising, bearing in mind  that 
there appears to be a close connection between singularity confinement and the Laurent phenomenon 
\cite{honepla,kanki_dkdv}, and the singularity confinement test was one of the first tools 
used to obtain discrete Painlev\'e equations \cite{grp}. On the other hand, it is known that 
singularity confinement is not sufficient for integrability \cite{hietvial}, and we have seen 
that nonintegrable examples can arise from Y-systems as well - indeed, integrability should be the 
exception rather than the rule. Nevertheless, our results show that cluster algebras 
are a promising source of new discrete integrable systems; and conversely, integrability 
provides powerful tools for the analysis of cluster algebras.  

In the future it would be interesting to connect this construction of discrete Painlev\'e equations to the 
geometrical approach pioneered by Sakai \cite{sakai}, which has subsequently  been extended to certain higher-order 
equations  \cite{kmnoy,tsuda}.  In this context, it has been noted in \cite{takenawa_tsuda} that  the 
Weyl group action on the tau-functions of some higher-order discrete Painlev\'e equations yields Laurent 
polynomials, which suggests a deeper connection with cluster algebras. 

\subsubsection*{Note added in proof:} As we were preparing the final version of this article for publication, 
we received a late report from one of the referees, which pointed out some closely related recent work by Okubo 
\cite{okubo}, of which we were previously unaware. Okubo also follows \cite{fordy_marsh}, considering both the Y-systems and 
the T-systems with coefficients (in the sense of \cite{nakanishi})  associated with certain periodic quivers; 
in particular he also points out the relation of 
the Somos-4 and Somos-5 Y-systems with q-Painlev\'e I and q-Painlev\'e II equations, respectively. 
Without having had the chance to make a thorough comparison of the latter work with our results, we note that 
the general T$_z$-system we consider has order $N+\tilde{r}$, as in Proposition \ref{tzsubs}, whereas  in \cite{okubo} 
each Y-system taken together with the associated T-system with coefficients has overall order $2N$. 
Preliminary examples suggest that there should be a gauge transformation linking  
(\ref{czrec}) to the corresponding T-system with coefficients, and we plan to study this in future work.

\subsubsection*{Acknowledgments:}  The authors are grateful for the comments by the referees, and would like to thank the organisers 
of the meeting on {\em Discrete Integrable Systems and Cluster Algebras} in the  Graduate School 
of Mathematical Sciences at the University of Tokyo,  where they first met in 2010.
ANWH is especially grateful to T. Nakanishi, who drew his attention to the discrepancy between 
the solutions of T-systems and Y-systems during the meeting on {\em Cluster Algebras and Statistical Physics} 
at ICERM, Brown University in 2011. 
RI is grateful for hospitality during her visit to the University of Kent  in 
October 2013.   
The work of RI is partially supported by JSPS KAKENHI Grant Number
26400037.

\section*{Appendix: Proof of Proposition \ref{palilem}}
To construct a palindromic basis for $\mathrm{im}\, B$, working over $\Q$ to begin with, one can start with 
any vector ${\bf v}\in \mathrm{im}\, B$. If ${\bf v}$ does 
not have palindromic support, then apply (positive or negative) powers 
of s to r$({\bf v})$ to obtain a vector s$^k(\mathrm{r}({\bf v}))$ 
for $k$ such that $\mathrm{supp}({\bf v})=\mathrm{supp}(\mathrm{s}^k(\mathrm{r}({\bf v})))$; 
then make the replacement  
${\bf v} \rightarrow  {\bf v}+ \mathrm{s}^k(\mathrm{r}({\bf v}))$, where the 
latter vector does have palindromic support. Now given 
the length $\ell = |\mathrm{supp}({\bf v})|$, if necessary one can 
apply s$^{-1}$ sufficiently many times, to replace 
${\bf v} \rightarrow \mathrm{s}^{-k^*}({\bf v})$, for $k^*\geq 0$ such that the latter 
vector has support $[1,\ell ]$. Due to Lemma~\ref{rsim}, 
having prepared ${\bf v}$ suitably, 
the set $S=\{ \mathrm{s}^j({\bf v}) \, | \, j\in [0,N-\ell ] \}$ 
is a palindromic basis for a subspace $V\subset \mathrm{im}\, B$, having dimension $N-\ell +1$,  

and if $N-r+1=\ell$ then $V$ coincides with $\mathrm{im}\, B$. Otherwise there 
is some row ${\bf b}_i$ of $B$ which is not in the span of the basis generated by 
${\bf v}$. It may be that $ |\mathrm{supp}({\bf b}_i)|<\ell$, but if not one can 
always subtract multiples of the vectors in $S$ to obtain a new vector ${\bf v}'$ 
whose support has length less than $\ell$, and by applying the same process 
as for ${\bf v}$ before one obtains a palindromic basis $S'$  
for a subspace   $V'\subset \mathrm{im}\, B$ of dimension larger than 
that of $V$. Repeating the same steps sufficiently many times 
produces a palindromic basis for $\mathrm{im}\, B$.  


To work over $\Z$, assume that a palindromic basis 
$\{ {\bf v}_1, {\bf v}_2, \ldots , {\bf v}_r \}$  
for $\mathrm{im}\, B$ is given; then without loss of generality 
(by suitable rescaling if necessary) 
$$
{\bf v}_1= (a_1^*, a_2^*,\ldots , a_{N-r+1}^*, 0,\ldots ,0),  
$$ 
where $a_j^*$ are integers (with $a_1^*\neq 0$ and $a_j^* = a_{N-r-j+2}^*$) 
such that the highest common factor of the non-zero entries is 1. 
Thus ${\bf v}_j\in \mathrm{im}\, B_\Z$ for $j\in [1,r]$, but it 
is not immediately clear that these vectors provide a $\Z$-basis. 
To see this, 
pick any  $\Z$-basis for $\mathrm{im}\, B_\Z$, and use it to form 
the rows of an $r\times N$ matrix $A=(a_{ij})$. By applying elementary 
row operations, equivalent to premultiplying by a unit matrix 
in  $\mathrm{Mat}_{r}(\Z)$, one can arrange it so that $A$ 
has the upper-triangular form 
$$ 
A = 
\left(\bear{cccccc} 
a_{11} & \cdots 
& \cdots  & \cdots & \cdots & a_{1N}  

                            \\ 
0     & a_{22}  & \cdots & \cdots & \cdots & a_{2N}  
                             \\ 

\vdots & \ddots & \ddots  & & & \vdots 
                              \\ 
0    &  \cdots   & 0  & a_{rr} & \cdots & a_{rN} 

\eear 
\right), 
$$ 
where $a_{jj}\neq 0$ for $j\in [1,r]$, and the highest common factor of the non-zero entries 
in each row is 1. Now consider the last row of $A$: this vector must have support 
$[r,N]$, since otherwise it would be a vector in  $\mathrm{im}\, B$ of length 
less than $N-r+1$, which would generate a palindromic basis for a space of 
dimension greater than $r$; and similarly, this vector must be a multiple of 
${\bf v}_r$, since otherwise one could subtract $(a_1^*)^{-1}a_{rr}{\bf v}_r$ 
from the last row to obtain a vector in $\mathrm{im}\, B$ having support of smaller 
length. Then since both ${\bf v}_r$ and the last row are both 
integer vectors whose non-zero entries have highest common factor 1, 
this last row must equal $\pm {\bf v}_r$; with out loss of generality take the plus sign. 
Proceeding by induction, assume that rows $j+1$ to $r$ of $A$ can be taken to be 
the vectors ${\bf v}_{j+1}, \ldots , {\bf v}_r$, and let ${\bf v}_j'$ 
denote row $j$. The vector ${\bf v}_j$ belongs to the span of rows 
$j$ to $N$ (which span the vectors with support in $[j,N]$), 
so it can be written as a $\Z$-linear combination 
$$ 
{\bf v}_j = c_j {\bf v}_j' + c_{j+1} {\bf v}_{j+1} + \ldots + c_N{\bf v}_N, 
$$    
with $c_j\neq 0$. The non-zero entries of the latter give a sequence of equations 
beginning 
$$ 
a_1^* = c_j a_{jj}, \quad  a_2^* = c_j a_{j,j+1}+ c_{j+1}a_1^*, 
\quad a_3^* = c_j a_{j,j+2}+ c_{j+1}a_2^*+ c_{j+2}a_1^*, \quad \ldots \, .
$$ 
If $p|c_j$ then the first equation in this sequence implies that 
$p|a_1^*$, whence the second equation gives $p|a_2^*$, and then the third 
implies $p|a_3^*$, and so on, meaning that $p$ is a common factor of the 
non-zero entries in ${\bf v}_1$. Hence $c_j = \pm 1$, and so 
${\bf v}_j$ can be taken to replace ${\bf v}_j'$ 
in row $j$ of $A$. By induction, the 
palindromic basis provides a $\Z$-basis for $\mathrm{im}\, B_\Z$, 
unique up to multiplying all 
basis vectors by $-1$. This completes the proof. 

\section*{References}


\begin{thebibliography}{99}

\bibitem{alman} 
J.~Alman, C.~Cuenca and J.~Huang, {\tt arXiv:1309.0751v2}

\bibitem{assem}I. Assem, C. Reutenauer and D. Smith,
Adv. Math. {\bf 225} (2010) 
3134--3165. 

\bibitem{bellon_viallet} 
M.P. Bellon and C.M. Viallet, 
Commun. Math. Phys. {\bf 204} (1999) 425--437. 


\bibitem{cima} 
A. Cima, A. Gasull and V. Ma\~{n}osa, 
Dynamical Systems {\bf 28}  (2013) 518--538. 

\bibitem{qsystems} 
P. Di Francesco and R. Kedem,
Lett. Math. Phys. {\bf 89} 
(2009) 183--216.

\bibitem{difrancesco} P. Di Francesco,
Lett. Math. Phys. {\bf 96} (2011) 299--324. 

\bibitem{QRT} J.J. Duistermaat, 
{\it Discrete Integrable Systems: QRT Maps and Elliptic Surfaces},
Springer (2010). 


\bibitem{eager}R. Eager, S. Franco and K. Schaeffer,
JHEP {\bf 06} (2012) 106. 

\bibitem{eager2} 
R. Eager and  S. Franco, 
JHEP {\bf 09} (2012) 038. 


\bibitem{fg}L. Faybusovich and M. Gekhtman,
J. Math. Phys. {\bf 41} (2000) 2905--2921.



\bibitem{fockgon}
V.V. Fock and  A.B. Goncharov,
Ann. Sci. \'Ec. Norm. Sup\'er. 
{\bf 42} (2009) 865--930.

\bibitem{fz1} S.~Fomin and A.~Zelevinsky,
J. Amer. Math. Soc. \textbf{15} (2002) 497--529.

\bibitem {fz} S.~Fomin and A.~Zelevinsky,
Adv. Appl. Math. \textbf{28} (2002)  119--144.

\bibitem{fziv}S.~Fomin and A.~Zelevinsky,
Compositio Mathematica {\bf 143} (2007) 112--164. 

\bibitem{fordy_marsh}A.P. Fordy and R.J. Marsh,
Journal of Algebraic Combinatorics,  \textbf{34} (2011) 19--66. 

\bibitem{fordy_rec} A.P. Fordy, 
Phil. Trans. Roy. Soc. A \textbf{369} (2011) 1264--1279

\bibitem{sigma}  A.P. Fordy and A.N.W. Hone, 
SIGMA 
{\bf 7} (2011) 091. 

\bibitem{fh} 
A.P.Fordy and A.N.W.Hone, Commun. Math. Phys. {\bf 325} (2014) 527--584; 
{\tt arXiv:1207.6072v1}  

\bibitem {gale} D.~Gale, 
Mathematical Intelligencer \textbf{13(1)} (1991),
40--42; 
Mathematical Intelligencer \textbf{13(4)} (1991),
49--50.

\bibitem{gsv}M. Gekhtman,  M. Shapiro and  A. Vainshtein,
Mosc. Math. J. {\bf 3} (2003) 899--934.

\bibitem{gsvduke}
M. Gekhtman,  M. Shapiro and  A. Vainshtein, 
Duke Math.~J.  {\bf 127} (2005) 291--311. 

\bibitem{gk}A.B. Goncharov and R. Kenyon,
Ann. Sci. \'Ec. Norm. Sup\'er. {\bf 46} (2013) 747--813. 

\bibitem{grp} 
B. Grammaticos, A. Ramani, V. Papageorgiou, 
Phys. Rev. Lett. {\bf 67} (1991) 1825.  

\bibitem{heidhogan08} 
P. Heideman and E. Hogan,
Elec. J. Comb. {\bf 15} (2008) $\#$R54.

\bibitem{hietvial} 
J. Hietarinta and C.M. Viallet,  
Phys. Rev. Lett. {\bf 81} (1998) 325. 


\bibitem{hir}R.~Hirota, 
Phys. Rev. Lett. {\bf 27} (1972) 1192--1194.

\bibitem {honeblms} A.N.W.~Hone,
Bull. Lond. Math. Soc. \textbf{37} (2005) 161--171.

\bibitem{hones5} A.N.W.~Hone,
Trans. Amer. Math. Soc.
\textbf{359} (2007) 5019--5034.

\bibitem{honepla} A.N.W. Hone,
Physics Letters A {\bf 361} (2007) 341--345. 

\bibitem{honelaur} A.N.W.~Hone.
SIGMA 
{\bf 3} (2007) 022. 

\bibitem{side6_poster}A.N.W.~Hone,
{\tt arXiv:0807.2538}

\bibitem{numberpoly} A.N.W.~Hone,
in J. McKee \& C. Smyth (eds.), \textit{Number Theory and Polynomials}, LMS Lect. Notes Series
{\bf 352}, Cambridge (2008) 
188--210.


\bibitem{honep} A.N.W. Hone, 
in A.V. Mikhailov (ed.), \textit{Integrability}, Lect. Notes Phys. {\bf 767}, 
Springer, Berlin-Heidelberg (2009) 
245--277.   

\bibitem{hones6} A.N.W. Hone,
Applicable Analysis {\bf 89} (2010) 473--492.


\bibitem{dkdv} 
A.N.W.~Hone, P.H.~van der Kamp, G.R.W.~Quispel 
and D.T.~Tran, Proc. Roy. Soc. A {\bf 469} (2013) 20120747. 

\bibitem{honeward}
A.N.W.~Hone and C. Ward, 
Bull. Lond. Math. Soc. {\bf 46} (2014) 503--516.   

\bibitem{iikns} 
R. Inoue, O. Iyama, A. Kuniba, T. Nakanishi, J. Suzuki, 
Nagoya Math. J. {\bf 197} (2010) 59--174.


\bibitem{inoue} 
R. Inoue and T. Nakanishi,
RIMS Kokyuroku Bessatsu {\bf B28} (2011) 63--88.  

\bibitem{kmnoy} 
K.Kajiwara, T.Masuda, M.Noumi, Y.Ohta and Y.Yamada, 
S\a'{e}min. Congr. {\bf 14} (2006) 169--198. 


\bibitem{kanki_dkdv} 
M. Kanki, J. Mada and T. Tokihiro, 
J. Phys. A: Math. Theor. {\bf 47} (2014) 065201.   

\bibitem{keller_scher}B. Keller and S. Scherotzke,  
Adv. Math. {\bf 228} (2011) 1842--1862. 
 
\bibitem{yuji_lauren}
Y. Kodama and L. Williams, 
Invent. math. (2014) 
{\tt doi:10.1007/s00222-014-0506-3} 


\bibitem{zabrodin} I. Krichever, O. Lipan, P. Wiegmann and A. Zabrodin, 
Commun. Math. Phys. {\bf 188} 
(1997) 267--304. 

\bibitem{asymmetric_dp} 
M.D. Kruskal, K.M. Tamizhmani, B. Grammaticos and A. Ramani, 
Regul. Chaotic Dyn. {\bf 5} (2000) 274--280. 

\bibitem{kuskly}V.B. Kuznetsov and E.K. Sklyanin, 
J. Phys. A: Math. Gen. {\bf 31} (1998) 2241.

\bibitem{lp}
T. Lam and P. Pylyavskyy, 
{\tt arXiv:1206.2611}



\bibitem{lobb}S. Lobb and F. Nijhoff, J. Phys. A {\bf 42} (2009) 454013. 

\bibitem{maeda}S. Maeda, 
Proc. Japan Acad. {\bf 63}, Ser. A (1987) 198--200. 
 

\bibitem{nakanishi}T. Nakanishi, 
in A. Skowronski \& 
K. Yamagata (eds.), {\it Representations of algebras and related topics}, 
EMS Series of Congress Reports, 
Eur. Math. Soc. (2011)  
407--444. 

\bibitem{okubo} N. Okubo, 
RIMS Kokyuroku Bessatsu {\bf B41} (2013) 25--41. 


\bibitem{qrt1} G.R.W.~Quispel, J.A.G.~Roberts and C.J.~Thompson,
Physics Letters A
\textbf{126} (1988) 419--421; 
Physica D \textbf{34} (1989) 183--192.

\bibitem{dpi}A. Ramani, B. Grammaticos and J. Satsuma, 
J. Phys. A: Math. Gen. {\bf 28} (1995) 4655--4665. 



\bibitem{ruijs}S.N.M. Ruijsenaars,
in 
B.A. Kupershmidt (ed.), 
\textit{Integrable and
Super-Integrable Systems}, World Scientific, Singapore (1990) 165--206.

\bibitem{sakai} 
H. Sakai, 
Commun. Math. Phys. {\bf 220} (2001) 165--229. 

\bibitem{scott} 
J.S. Scott, 
Proc. Lond. Math. Soc. {\bf 92} 
(2006) 345--380. 


\bibitem{tsuda_elliptic} 
T. Tsuda,  
J. Phys. A: Math. Gen. {\bf 37} (2004) 2721. 

\bibitem{tsuda} 
T. Tsuda,  
Lett. Math. Phys. {\bf 77} (2006) 21--30. 

\bibitem{takenawa_tsuda} 
T. Tsuda and T. Takenawa, 
Adv. Math. {\bf 221} (2009) 936--954. 

\bibitem{veselov} A.P. Veselov, 
Russ. Math. Surveys \textbf{46} (1991) 1--51.


\bibitem{zam}Al. B. Zamolodchikov, 
Phys. Lett. B \textbf{253} (1991) 391--394. 
 
\end{thebibliography}
\end{document}